\documentclass[manuscript, screen, acmsmall]{acmart}
\AtBeginDocument{%
  }

\setcopyright{acmlicensed}
\copyrightyear{2026}
\acmYear{2026}
\acmDOI{XXXXXXX.XXXXXXX}

\acmJournal{HEALTH}
\acmVolume{1}
\acmNumber{1}
\acmArticle{1}
\acmMonth{1}
\usepackage{tikz}
\usetikzlibrary{shapes.geometric}
\usetikzlibrary{shapes.multipart} 
\usetikzlibrary{positioning,decorations.pathreplacing}
\usepackage{dirtree}
\usepackage{pgfplots}
\usepgfplotslibrary{groupplots}
\pgfplotsset{compat=1.17}
\usetikzlibrary{pgfplots.statistics}

\usepackage[most]{tcolorbox}

\usepackage{graphicx} 

\usepackage[capitalize]{cleveref}
\usepackage{listings}
\usepackage{xcolor}

\usepackage{todonotes}
\usepackage{fontawesome}
\usepackage{wrapfig}
\usepackage{opensans}
\newboolean{SHOWCOMMENTS}
\setboolean{SHOWCOMMENTS}{true}
\ifthenelse{\boolean{SHOWCOMMENTS}}{
\newcommand{\tdMK}[1]
{\todo[color=blue!25,inline]{\footnotesize{\textbf{ Martin:}} #1}}
}{
\newcommand{\tdMK}[1]{}
}

\newcommand{\vismode}{anonymized}
\newcommand{\reviewanonymized}[2]{%
  \ifx\vismode\anonymized
    #1%
  \else
    #2%
  \fi
}

\setcopyright{none}
\begin{document}
\lstset{
	language=, 				        
	basicstyle=\scriptsize\ttfamily, 	
	keywordstyle=\color{red}\bfseries,	
	identifierstyle=\color{black},		
	commentstyle=\color{DarkGreen},		
	stringstyle=\ttfamily\color{blue}, 	
	breaklines=true, 			        
	numbers=none, 				        
	numberstyle=\tiny\color{gray}, 		
    numbersep=-25pt,              
	frame=tb, 				            
	backgroundcolor=\color{gray!10}, 	
	tabsize=2,				            
    captionpos=b                        
}
\renewcommand\lstlistingname{Code Snippet}
\renewcommand\lstlistlistingname{Code Snippets}
\title[Semi-automated interface generation for efficient simulation backends]{Manual binding and semi-automated interface generation for efficient simulation backends with user-friendly frontends: A~demonstration for epidemic use cases}

\author{Maximilian Betz}
\authornote{Both authors contributed equally to this research.}
\email{m.betz@fz-juelich.de}
\orcid{0009-0001-9181-9999}
\affiliation{%
  \institution{ICE-1, Institute of Climate and Energy Systems, Forschungszentrum Jülich}
  \city{Jülich}
  \country{Germany}
}
\author{Daniel-Michel Richter}
\orcid{0009-0000-9741-6050}
\authornotemark[1]
\email{daniel-michel.richter@dlr.de}
\affiliation{%
  \institution{Institute of Software Technology, German Aerospace Center}
  \city{Cologne}
  \country{Germany}
}

\author{Daniel Abele}
\orcid{0000-0001-7021-1573}
\affiliation{%
  \institution{Institute of Software Technology, German Aerospace Center}
  \city{Cologne}
  \country{Germany}
}



\author{René Schmieding}
\orcid{0000-0002-2769-0270}
\affiliation{%
  \institution{Institute of Software Technology, German Aerospace Center}
  \city{Cologne}
  \country{Germany}
}

\author{Henrik Zunker}
\affiliation{%
  \institution{Institute of Software Technology, German Aerospace Center}
  \city{Cologne}
  \country{Germany}
}
\email{henrik.zunker@dlr.de}
\orcid{0000-0002-9825-365X}

\author{Martin J. Kühn}
\affiliation{%
  \institution{Institute of Software Technology, German Aerospace Center, Cologne}
  \institution{and Bonn Center for Mathematical Life Sciences and Life and Medical Sciences Institute, University of Bonn, Bonn}
  \country{Germany}
}
\orcid{0000-0002-0906-6984}
\email{martin.kuehn@dlr.de}

\renewcommand{\shortauthors}{Betz et al.}

\begin{abstract}
Challenges and chances of mathematical software are directly entangled with the programming language used. Depending on the background of developers and users, conflicts between different goals may arise. While several, e.g., scripting languages allow a low barrier and user-friendly entry to mathematical modeling and simulation, corresponding implementations will most likely not allow timely outcomes for detailed simulations. 

In infectious disease modeling, while several large-scale software frameworks rely on C or C++ implementations, most models are implemented in Python or R. While the latter allow for fast prototyping and more interactive development, it has been shown for several applications that C or C++ implementations outperform plain Python or R implementations by factors between one and three orders of magnitude. In particular, in times of pandemics, the urgent need for swift estimation of the system's dynamics makes the use of large-scale models in these noncompiled languages prohibitive.

In order to create synergies and close the gap between scientists from the application and the computational domain, binding libraries allow to execute entire, compiled software frameworks from languages such as Python. However, with iterative model development, extension, or addition of new models, implementing bindings can represent another relevant barrier to developers of the C or C++ backends. In this paper, we show how a large-scale software framework with a multitude of infectious disease models can be efficiently exposed to Python. In addition, we suggest a semi-automated way to generate interface code by analyzing generic C++ code through abstract syntax trees, enabling a modularized approach working with intermediate representations.  
\end{abstract}




\maketitle

\section{Introduction}

With individual programming languages serving different purposes, different challenges and opportunities arise. In particular, developers and users need to gain substantially different levels of computational expertise for implementing and using mathematical methods. In epidemiological modeling, languages such as R and Python are most widely spread~\cite{bicker2026memilio}, as they offer a low-barrier access to mathematical-epidemiological modeling. However, it has been shown for different applications~\cite{pereira_energy_2017,portegies_zwart_ecological_2020} that compiled languages such as C or C++ outperform Python with respect to time to solution by factors around 50, 100, or more. This well-known fact has led to the development of C, C++, or Fortran compiled packages to, e.g., solve ordinary differential equation-based (ODEs) models in R or Python~\cite{desolve,2020SciPy-NMeth} and for the well known pytorch library, it was recently stated, how non-Python environments such as C++ help to bridge the gap between research prototypes and production deployment~\cite{stryker_ibm_torchscript}. However, addressing individual software components such as numerical ODE solvers for ODE-based problems might not be enough if other costly tasks are conducted in a scripting approach. In~\cite{bicker2026memilio}, it has recently been shown that ODE-based problem formulations with a modest number of spatially distributed subpopulations are simulated 1-2 orders of magnitude faster in fully-bound C++ than when only seeking Fortran or C compiled routines to solve the ODE system. With computation times at the order of seconds, individual simulations will rarely present a bottleneck. Nevertheless, with iterative model development or thousands of simulations necessary for parameter exploration and model calibration, execution will become incompatible with the high timely pressure that is imposed by unfolding or ongoing epidemics. In that direction, efficiently implemented models contribute substantially to timely outputs -- an important use factor that increases modeling relevance for policy~\cite{Roebl2026}.

While several large-scale software models for infectious disease modeling like OpenABM~\cite{OpenABM}, EpiHiper~\cite{chen_epihiperhigh_2025}, FRED~\cite{grefenstette_fred_2013} are implemented in C and C++, primarily targeting users with background knowledge in these languages, others like Covasim~\cite{kerr_covasim_2021} or MatSim-episim~\cite{mueller2021episim} provide optimized Python or Java implementations to increase the developer and user base. However, eventually, these optimized versions will stay behind the potential of software written in compiled languages. Even other frameworks such as Repast~\cite{collier_parallel_2013,collierDistributedAgentBased2022} provide separated repositories with models implemented in either C++ or Python. BoostPython~\cite{boost_python}, pybind11~\cite{pybind}, and nanobind~\cite{nanobind} are well established methods to provide Python bindings to C++ implementations. More recent developments based on the Cling C++ interpreter promise better Python binding performance than established solutions with pybind11 but are also less popularized yet~\cite{kundu_efficient_2023}. The software frameworks EMOD~\cite{bershteyn_implementation_2018}, EpiABM~\cite{gallagher_epidemiological_2022}, and MEmilio~\cite{bicker2026memilio} provide Python frontends to their C/C++ backends where the latter both use pybind11. Through this approach, EMOD, EpiABM, and MEmilio are capable to offer low-barrier interfaces tailored to domain-specific users. It has to be noted that in the related domain of veterinary diseases, the provision of R, Python, or even Javascript frontends to compiled language implementations has an already longer history as, e.g., seen by SimInf~\cite{widgren2019siminf} or BICI~\cite{the_iteam_BICI}. However, the provision of these interfaces comes at substantial cost for developers that need to implement a substantial number of additional lines of code. In these, additional difficulties arise as developers need to understand the particularities of different programming languages using objects through, e.g., pointers, references, or different types of copies. On the other hand, binding large code bases yields a lot of opportunities and, compared to fully new implementations, comes at substantially reduced cost.

To avoid or reduce the cost of manual binding implementations, (semi-)automated approaches have been proposed. AutoWIG~\cite{fernique_autowig_2017} or CppInterOp~\cite{jomy_cppinterop_2025} are two approaches that have been suggested for the automated binding generation of C++ libraries. For our application, we distinguish two complementary use cases where automation can come into play. In cases where a clearly separated software module (e.g., a new disease model) is added to the software, bindings can be directly generated with the backend code. Here, a unified description as in MEmilio~\cite{bicker2026memilio} or a domain-specific language (DSL) as in EMULSION~\cite{picault_emulsion_2019} is needed. A popular example of a DSL in biological applications is the Systems Biology Markup Language (SBML) standard~\cite{hucka_systems_2003} which facilitates the task of generating model backend and interface simultaneously from a description in xml or yaml formats. DSLs are also used in other domains, e.g., to generate stencil-based code for high-performance computing applications~\cite{schmitt_exaslang_2014,bauer_lbmpy_2021,alhaddad_highpermeshes_2022}. In cases where the backend code is already available in the compiled language, an analysis of the abstract syntax tree (AST) can enable the semi-automated generation of binding code. For our application, we furthermore distinguish the second use case by the settings where an existing but entire module might need a binding interface or where bindings should only be generated for small entities such as individual functions. 

The paper is structured as follows. In~\cref{sec:modular}, we first briefly revisit the descriptions of available models in the MEmilio software framework and discuss the advantages and necessity of modular and generic implementations before, in~\cref{sec:pymio}, explaining the goals, challenges, and essentials chosen in MEmilio to provide C++ implementations to Python. In~\cref{sec:pygen}, we explain more deeply the (semi-automated) development of Python bindings for existing and new models sharing the same or similar descriptions by analyzing ASTs and transforming them to intermediate representations before generating the target code. We present a modularized approach based on handlers and scanners for ASTs that are connected through an intermediate representation with a code generator for different applications. In~\cref{sec:results}, we show results of the semi-automated generation with respect to code completeness of a comparable MEmilio model and evaluate the performance achieved through Python before, eventually, drawing a conclusion in~\cref{sec:conclusion}.

\section{Modular and generic implementations for infectious disease models}
\label{sec:modular}

Manual binding between programming languages is cumbersome work. Wherever modularization and generic implementations avoid (partial) code redundancy across the framework backend, the overhead of manual code binding can be substantially reduced. When it comes to a (semi-)automated generation of binding code, unified patterns are prerequisites for a functional automatization. 

The MEmilio software framework~\cite{bicker2026memilio} considered as a base for this article is already set up in a modular fashion and patterns have been unified across different modules of the framework. To better understand the manual and (semi-)automated bindings on top of the C++ backend, we here present structures and several examples from models within the MEmilio framework.

In Code Snippet~\ref{code:infstates}, we provide the flexible definition of infection states for the example of an SEIR-type model representing four infection states (Susceptible, Exposed, Infected, Recovered) and always also integrating a count variable representing the number of states, used for iterating over the states in various places. All models define their individual namespaces inside the global \texttt{mio} namespace. In Code Snippet~\ref{code:params}, we show the struct-based definition of a parameter with name \texttt{TimeExposed} for an equation-based model with age stratification. The parameter is templated for the floating point type \texttt{FP} which allows distinction between double precision or the simultaneous consideration of the functional value and an active automatic differentiation (AD) type~\cite{Naumann2011}. For the latter case, relevant components of MEmilio are overloaded for a wide range of equation-based models such that derivative information is directly propagated together with the functional value of the variable. Theoretically, the \texttt{FP} type could also be used to, e.g., use single precision. For models considering aggregated population counts such as in equation-based models, an \texttt{UncertainValue} is used as a generalization of a floating point value, with the allowance of (re)sampling the value from a given distribution. For more complex, agent-based models (ABMs)~\cite{KERKMANN_ABM_2025,korf_effect_2026}, the \texttt{UncertainValue} in \texttt{CustomIndexArray} is generalized to an \texttt{AbstractParameterDistribution} and stratified according to a \texttt{VirusVariant}.

\vspace{0.5cm}
\begin{minipage}[c]{0.48\textwidth}
\begin{lstlisting}[caption={\textbf{Enumeration-based definition of infection states in SEIR model.}},label={code:infstates}]
namespace oseir
{
enum class InfectionState
{
    Susceptible,
    Exposed,
    Infected,
    Recovered,
    Count
};
} 
\end{lstlisting}
\end{minipage}
\hfill
\begin{minipage}[c]{0.48\textwidth}
\begin{lstlisting}[caption={\textbf{Struct-based definition of parameter in equation-based model with age stratification and Exposed state.}},label={code:params}]
template <typename FP>
struct TimeExposed {
    using Type = CustomIndexArray<UncertainValue<FP>, AgeGroup>;
    static Type get_default(AgeGroup size)
    {
        return Type(size, 5.2);
    }
    static std::string name()
    {
        return "TimeExposed";
    }
};
\end{lstlisting}
\end{minipage}

While the definition of infection states is identical across all models, transitions or \textit{flows} between infection states are particularly defined for ODE-based models, from which more than 10 different models are implemented in MEmilio. In Code Snippet~\ref{code:flows}, we present the flows and model definition for the ODE-based SEIR-type model. Here, only three forward flows from the prior to the later state (without branching nor backward flows) are possible. ODE-based models derive from a \texttt{FlowModel} integrating the floating point type, the list of infection states, a correspondingly resolved population and the list of parameters and flows.

\vspace{0.5cm}
\begin{minipage}[c]{0.96\textwidth}
\begin{lstlisting}[caption={\textbf{List-based definition of flows between infection states and model constructor for an ODE-based SEIR model.}},label={code:flows}]
namespace oseir
{
using Flows = TypeList<Flow<InfectionState::Susceptible, InfectionState::Exposed>,
                       Flow<InfectionState::Exposed,     InfectionState::Infected>,
                       Flow<InfectionState::Infected,    InfectionState::Recovered>>;
template <typename FP>
class Model
    : public FlowModel<FP, InfectionState, mio::Populations<FP, AgeGroup, InfectionState>, Parameters<FP>, Flows>
{
[...]
\end{lstlisting}
\end{minipage}

The correspondingly resolved population is given by a \texttt{Populations} class which stores the numbers of individuals according to different features, e.g., age and infection state in an array, a \texttt{CustomIndexArray}. The \texttt{Populations} object uses a variadic template to allow stratification of the population by an arbitrary number of features or \texttt{Categories} as shown in Code Snippet~\ref{code:populations}. Stratification by age is resolved through an \texttt{AgeGroup} struct using the curiously recurring template pattern (CRTP). As shown in Code Snippet~\ref{code:age}, \texttt{AgeGroup} is used as a tag type for the typesafe index \texttt{Index}, which we refrain from detailing for its technical implementation.

\vspace{0.5cm}
\begin{minipage}[c]{0.96\textwidth}
\begin{lstlisting}[caption={\textbf{Definition of populations stratified according to a variadic number of categories.}},label={code:populations}]
template <typename FP, class... Categories>
class Populations : public CustomIndexArray<UncertainValue<FP>, Categories...>
[...]
\end{lstlisting}
\end{minipage}

\vspace{0.5cm}
\begin{minipage}[c]{0.96\textwidth}
\begin{lstlisting}[caption={\textbf{AgeGroup struct to allow for model stratification by age.}},label={code:age}]
struct AgeGroup : public Index<AgeGroup> {
    AgeGroup(size_t val)
        : Index<AgeGroup>(val)
    {    }
[...]
\end{lstlisting}
\end{minipage}

The remaining \texttt{Parameters} from the model definition are derived from a \texttt{ParameterSet} which represents a rather technical definition of a parameter set defined at compile time and where parameters are added as template parameters (tags). The ParameterSet constructor exists if all parameters are default constructible and the parameters are passed with their names as shown in Code Snippet~\ref{code:parameterlist}.

\vspace{0.5cm}
\begin{minipage}[c]{0.96\textwidth}
\begin{lstlisting}[caption={\textbf{Definition of the parameter list for an ODE-based SEIR model.}},label={code:parameterlist}]
template <typename FP>
using ParametersBase =
    ParameterSet<TransmissionProbabilityOnContact<FP>, TimeExposed<FP>, TimeInfected<FP>, ContactPatterns<FP>>;

template <typename FP>
class Parameters : public ParametersBase<FP>
{
public:
    Parameters(AgeGroup num_agegroups)
        : ParametersBase<FP>(num_agegroups)
        , m_num_groups{num_agegroups}
    {
    }
[...]
\end{lstlisting}
\end{minipage}

The prior structures are reused throughout the MEmilio framework to realize a large range of mathematical-epidemiological models. For instance, any age-resolved model as presented in Code Snippet~\ref{code:flows}, can be created on compile time with $n$ age groups by simply calling the constructor in the model's name space with \texttt{Model($n$)}. 

We refrain from providing more technical details or generalizations as, e.g., used in ABMs and refer to the code base and documentation of MEmilio\footnote{\url{https://memilio.readthedocs.io}}.

\section{From C++ backend to a practitioner-targeted Python interface}
\label{sec:pymio}

There are different levels of granularity on how to expose (parts of) a C++ library to Python. In the simplest case, we can bind particular simulation executables with command line interfaces (CLIs) in Python. While this approach is straightforward and requires minimal effort, it also substantially limits the users' interactions with the code backend. The extensive implementation of bindings is tedious and exhausting but can, on the other hand, become the more relevant use case. With flexible bindings, individual users obtain much more freedom in setting up a model according to their needs.

Once the decision for a flexible interface is taken, a comprehensive understanding of, both, programming languages and the architecture of the underlying backend software is required. Developers must be familiar with C++ concepts such as ownership semantics, references, object lifetimes, and template programming while simultaneously understanding how these concepts should be represented in a Pythonic way. In addition, they need to understand the internal structure of the backend to decide which classes, functions, and data structures form part of the public API.

\subsection{Development of the MEmilio Python interface}

As shown in~\cite{bicker2026memilio}, most infectious disease models are implemented in Python or R which means that the target user base of MEmilio is largely made up from Python developers. In order to allow these users a maximum of flexibility, the Python interface is provided as a fully functional module where users can adapt most of the libraries functionality. Allowing users to adapt parameters and scenarios also enables the coupling of established Python packages, e.g., for sensitivity analysis or parameter estimation. In contrast to a simple binding of a C++ executable with a CLI, our approach targets for a low-barrier and comprehensible line-by-line execution in Python. 

MEmilio's Python interface is made up by manually created bindings and bindings that can be generated semi-automatically as described in the next chapter. Both approaches have different, yet overlapping use cases. While the MEmilio ABM~\cite{KERKMANN_ABM_2025}, as a unique and flexible model, is developed and extended through changes and patches of limited size, generalizations of equation-based models might require extensive reimplementations based on recurring patterns. This means that MEmilio's Python interface provides access to the ABM and existing equation-based models through pre-implemented bindings, while the semi-automated generation of new bindings addresses new (equation-based) models and small-scale extensions of general C++ structures. In particular for new models of similar structures, the semi-automated generation maximizes its impact by substantially reducing implementation overhead. Building upon the structure of the C++ library, we provide one module with commonly used utils and submodules for each model where the interface for each module is defined in a single file.

The modular structure of MEmilio allows to create and adapt C++ models and simulations with a lot of flexibility. The Python interface strives to define a common interface between C++ and Python, making the functionality usable in a similar fashion on both ends allowing users to modify domain specific features like age group stratification. As a user, running simulations commonly only involves using a simulation function, which provides an easy interface for solving the model problem until a time point and automatically receiving the result as the return value. When a more granular utilization is needed, the \texttt{Simulation} class can be used directly. In Code Snippet~\ref{code:simulate}, we exemplarily show how the simulation functions using either compartmental values (\texttt{simulate}) or flows (\texttt{simulate\_flows}) are translated into Python.

A particular issue originates from templated functions which are resolved in C++ on compile-time but which cannot be handled on Python side in the same fashion. Instead of binding templated functions for each possibly relevant call, we differentiate between the full C++ functionality and the interface which provides the most relevant functionality only. As for instance shown in Code Snippet~\ref{code:simulate}, we only provide the simulation functions for double precision, while, theoretically, the C++ backend allows for substituting the data type. Similarly, we do not bind technical helper functions and low-level numerical functionality which can only be changed in C++ directly.

\vspace{0.5cm}
\begin{minipage}[c]{0.96\textwidth}
\begin{lstlisting}[caption={\textbf{Definition of the interface for simulations of an ODE model.}},label={code:simulate}]
m.def("simulate", &mio::simulate<double, mio::oseir::Model<double>>, "Simulates an ODE SEIR from t0 to tmax.", py::arg("t0"), py::arg("tmax"), py::arg("dt"), py::arg("model"), py::arg("integrator") = py::none());

m.def("simulate_flows", &mio::simulate_flows<double, mio::oseir::Model<double>>, "Simulates an ODE SEIR with flows from t0 to tmax.", py::arg("t0"), py::arg("tmax"), py::arg("dt"), py::arg("model"), py::arg("integrator") = py::none());
\end{lstlisting}
\end{minipage}

In Code Snippet~\ref{code:simulatecpppy}, we show how the simulate function for an ODE-SEIR model can now be called in particular C++ and Python examples. While we see slight differences, as e.g., the template parameter which is resolved on compile time in C++ and the namespace which is implicitly resolved in Python through our folder structure. For the C++ call, the last argument \texttt{integrator\_core} is not given and by default set to \texttt{nullptr} and likewise, Python sets the \texttt{integrator} to \texttt{py::none()}, which for both cases means that the default Runge-Kutta Cash-Karp5(4) integrator is chosen.

\vspace{0.5cm}
\begin{minipage}[c]{0.96\textwidth}
\begin{lstlisting}[caption={\textbf{Comparison of calling the C++ simulate routine for an ODE-SEIR model versus calling the bound Python routine.}},label={code:simulatecpppy}]
# C++
auto result = mio::simulate<FP>(t0, tmax, dt, model);
# Python
result = simulate(t0, tmax, dt, model)
\end{lstlisting}
\end{minipage}

Aside from model-specific bindings, the MEmilio Python bindings contain general binding functionality and utilize templated binding helper functions such as \texttt{bind\_population} shown in Code Snippet~\ref{code:population}. These functions improve the readibility of our interface and template deduction from the specific models then enables the corresponding model-specific binding. For instance with the variadic template \texttt{Cats}, the categoric stratification of the population along, e.g., infection states and age groups is inferred.

\vspace{0.5cm}
\begin{minipage}[c]{0.96\textwidth}
\begin{lstlisting}[caption={\textbf{Bindings for the templated binding helper function for connecting with the model population.}},label={code:population}]
template <class... Cats>
void bind_Population(pybind11::module_& m, std::string const& name,
                     mio::Tag<mio::Populations<double, Cats...>> /*tags*/)
{
[...]
\end{lstlisting}
\end{minipage}

\vspace{0.5cm}
\begin{minipage}[c]{0.96\textwidth}
\begin{lstlisting}[caption={\textbf{Binding of the TimeSeries object with consideration of pickling and conversion to numpy array.}},label={code:timeseries}]
void bind_time_series(py::module_& m, std::string const& name)
{
    bind_class<mio::TimeSeries<double>, EnablePickling::Required>(m, name.c_str())
        [...]
        .def(
            "as_ndarray",
            [](mio::TimeSeries<double>& self) {
                auto m = Eigen::Map<mio::TimeSeries<double>::Matrix>(self.data(), self.get_num_rows(), self.get_num_time_points());
                return Eigen::Ref<mio::TimeSeries<double>::Matrix>(m);
            },
            py::return_value_policy::reference_internal)
\end{lstlisting}
\end{minipage}

To appropriately target the Python user base, we intend to translate the library in a Pythonic way instead of creating a direct copy of the C++ backend. While following the general design of the C++ library, partial adaptations are advised to incorporate Pythonic coding conventions. By these adaptations, we intend to increase the usability and provide a familiarity for developers with Python experience. For instance, we always provide argument names that allow Python developers to hand over objects and parameters not only in the predefined C++ order but also allowing argument names that assist developers in setting up simulations. The \texttt{TimeSeries} object is used in MEmilio's backend for multidimensional, time-dependent arrays, in particular as the output of model simulations. When coupling MEmilio with other (Python) packages, e.g., for plotting with matplotlib, the TimeSeries output should be transformable to standardized objects. By enabling a conversion to numpy array, see Code Snippet~\ref{code:timeseries}, we facilitate the analysis of simulation outcomes in standardized post-processing or plotting scripts.

Running two languages side by side requires aligning their different memory management models. C++ focuses on manual memory management and RAII (Resource Acquisition Is Initialization), while Python has an automatic approach based on garbage collection and reference counting. Therefore, when introducing bindings, a central concern is the ownership and lifetime of objects shared across the language boundary.

Pybind11 addresses this with return value policies, which define how ownership of a C++ object is handled when it is exposed to Python. Deciding the correct policy can be hard for nontrivial types and depends on the design of the function. The wrong policy can lead to crashes or undefined behavior. For example, if Python wrongfully assumes ownership it may delete an object, even though the object is still used in C++. Or C++ may destroy an object that Python still references, as could happen with the function shown in Code Snippet~\cref{code:timeseries}, where the returned object references data that should be managed by C++. We use the \texttt{reference\_internal} policy, which leaves ownership with C++ but additionally keeps alive the parent object as long as the reference exists in Python. This prevents an early deallocation of the C++ object, which could otherwise fail silently and possibly lead to data corruption.

Eventually, when implementing additional code in Python, software testing has to be addressed. While generally, good software engineering practices advise to implement extensive unit testing, binding C++ libraries profits from an existing code base that ensures robust solutions without extensive (re)testing. In order to ensure high software quality at a reduced number of new tests and simultaneously minimize errors of interfaces, we opt for more high-level software or integration test on the interface side and compare outcomes to identical tests for the C++ backend. While this leaves a marginally higher chance for introducing errors on the Python interface side, we conclude that with the essential high-level functionality also the individual units behave the same for both interfaces.

\subsection{Challenges for new models and structures}

The primary design goal of the MEmilio Python interface is to provide a flexible and intuitive interface that evolves together with the MEmilio backend. As new epidemiological models and simulation features are continuously added, the Python bindings must be extended accordingly while maintaining a consistent user experience across all supported models.

The example of the TimeSeries binding from the previous section shows the importance of manual interaction when writing the interface: bindings may follow a common pattern but the decision of and how strongly adapting it to Pythonic coding conventions often relies on case to case decisions. However, the continuous evolution of MEmilio results in a continuous and potentially increasing overhead that cannot be addressed by manually written bindings as the framework grows. Every modification of the underlying C++ library potentially requires corresponding adaptations in the bindings. Maintaining the interface becomes an ongoing software engineering task that overly accompanies the further development of the simulation framework. 

\section{Semi-automated interface generation through ASTs of existing code}
\label{sec:pygen}

The arrival of publicly available coding agents has facilitated implementation tasks. However, vibe- or co-coded implementations might suffer from different types of issues such as errors from overconfident outputs, overly complex structures, and missing verification and validation~\cite{pmlr-v216-radev23a,spiess_correctnessLLMs,NEURIPS2023_43e9d647}. For our dedicated setting, classic code generation through AST analysis is better suited and, in addition, only requires local resources. Building on top of clear modularization, unified structures, and recurring patterns as introduced in~\cref{sec:modular}, we suggest a semi-automated approach for interface generation. Our approach is based on the compilation of several targets of the underlying software framework and can then be executed in a matter of seconds to minutes on laptop infrastructure. By providing different templates of the target code, we allow a model- or more general structure-based generation of the binding code. The resulting structure substantially accelerates the manual process of implementation by also avoiding typical typo or copy-and-paste errors. Our long term goal is to provide an automatic creation interface for new user-defined models, including their (full) exposure to Python without any prior knowledge needed of binding libraries. This goal is especially important for Python users, as both the base models, as well as the bindings are written in C++.

\subsection{Concept and workflow}\label{sec:concept_and_workflow}

\begin{figure}[!h]
\tikzset{
  block/.style = {rectangle, draw=black, fill=blue!20, 
                  text width=5em, text centered, rounded corners, minimum height=3em},
  arrow/.style = {thick,->,>=stealth}
}
\tikzstyle{data} = [trapezium, trapezium left angle = 70, trapezium right angle = 110, text centered, text width = 3cm, minimum height = 1cm, draw=black, fill=blue!10, trapezium stretches=false]
\tikzstyle{logic} = [rectangle, rounded corners, text centered, text width = 3cm, minimum width = 4cm, minimum height = 2.5cm, draw=black, fill=blue!30]
	\centering
	\begin{tikzpicture}[every node/.style={scale=0.55, font=\Large}]
		\node (ASTHandler) [logic] {\textbf{ASTHandler}};
            \node (Scanner) [logic, below of=ASTHandler, yshift=-8cm] {\textbf{Scanner}};
		\node (IR) [data, below of=Scanner, yshift=-3cm] {
		\vphantom{\begin{tabular}{c}w\end{tabular}} \\ 
		Intermediate\\ Representation  \\ \vphantom{\begin{tabular}{c}w\end{tabular}}};
		\node (Generator) [logic, below of=IR, yshift=-3cm] {\textbf{Generator}};
		
		\node (Sourcecode) [data, above of=ASTHandler, yshift=3cm, xshift=-5cm, fill=orange!30] {\vphantom{\begin{tabular}{c}z\end{tabular}} \\ C++ Target 0 \\ \vphantom{}};
            \node (Sourcecode1) [data, above of=ASTHandler, yshift=3cm, fill=orange!30] {\vphantom{\begin{tabular}{c}z\end{tabular}} \\ C++ Target ... \\ \vphantom{}};
            \node (Sourcecode2) [data, above of=ASTHandler, yshift=3cm, xshift=5cm, fill=orange!30] {\vphantom{\begin{tabular}{c}z\end{tabular}} \\ C++ Target N \\ \vphantom{}};
            \node (AST0) [data, below of=ASTHandler, yshift=-3cm, xshift=-5cm] {\vphantom{\begin{tabular}{c}z\end{tabular}} \\ AST 0\\ \vphantom{}};
            \node (AST) [data, below of=ASTHandler, yshift=-3cm] {\vphantom{\begin{tabular}{c}z\end{tabular}} \\ AST ... \\ \vphantom{}};
            \node (ASTN) [data, below of=ASTHandler, yshift=-3cm, xshift=5cm] {\vphantom{\begin{tabular}{c}z\end{tabular}} \\ AST N \\ \vphantom{}};
        
		\node (Configuration) [data, below of=ASTHandler, xshift=10cm, yshift=-3cm] {Configuration (JSON-file and Python-class)};
        \node (User) [data, left of=Scanner, xshift=-6cm, yshift=1.5cm, fill=orange!30] {User-defined selection of structures};
		\node (Template) [data, right of=Generator, xshift=6cm, yshift=1.5cm] {\vphantom{\begin{tabular}{c}z\end{tabular}} \\ Template of the target code \\ \vphantom{}};
		\node (Binding) [data, below of=Generator, yshift=-3cm, fill=orange!30] {Target code with the C++-Python bindings};
		
	\draw[thick,<-] (ASTHandler) -- node[anchor=east] {Input} (Sourcecode);
        \draw[thick,<-] (ASTHandler) -- node[anchor=east] {Input} (Sourcecode1);
        \draw[thick,<-] (ASTHandler) -- node[anchor=east] {Input} (Sourcecode2);

        \draw[thick,<-] (AST0) -- node[anchor=east] {Create} (ASTHandler);
        \draw[thick,<-] (AST) -- node[anchor=east] {Create} (ASTHandler);
        \draw[thick,<-] (ASTN) -- node[anchor=east] {Create} (ASTHandler);
        
        \draw[thick,<-] (Scanner) -- node[anchor=east] {Transfer} (AST0);
        \draw[thick,<-] (Scanner) -- node[anchor=east] {Transfer} (AST);
        \draw[thick,<-] (Scanner) -- node[anchor=east] {Transfer} (ASTN);
		\draw[thick,<-] (Scanner) -| node[anchor=north] {} (Configuration);
        \draw[thick,<-] (ASTHandler) -| node[anchor=north] {} (Configuration);
        \draw[thick,<-] (Scanner) -| node[anchor=north] {} (User);
		\draw[thick,->] (Scanner) -- node[anchor=east] {Write} (IR);
		\draw[thick,<-] (Generator) -- node[anchor=east] {Transfer} (IR);
		\draw[thick,<-] (Generator) -| node[anchor=north] {} (Template);
		\draw[thick,->] (Generator) -- node[anchor=east] {Output} (Binding);
	\end{tikzpicture}
    \caption{\textbf{Workflow of the modular analysis and code generation process.} The blue boxes represent components of the modular pipeline while the orange boxes indicate input and output elements. Trapezoidal boxes denote data structures and rounded boxes represent logic classes or active components.}
    \label{fig:pygen_workflow_new}
\end{figure}
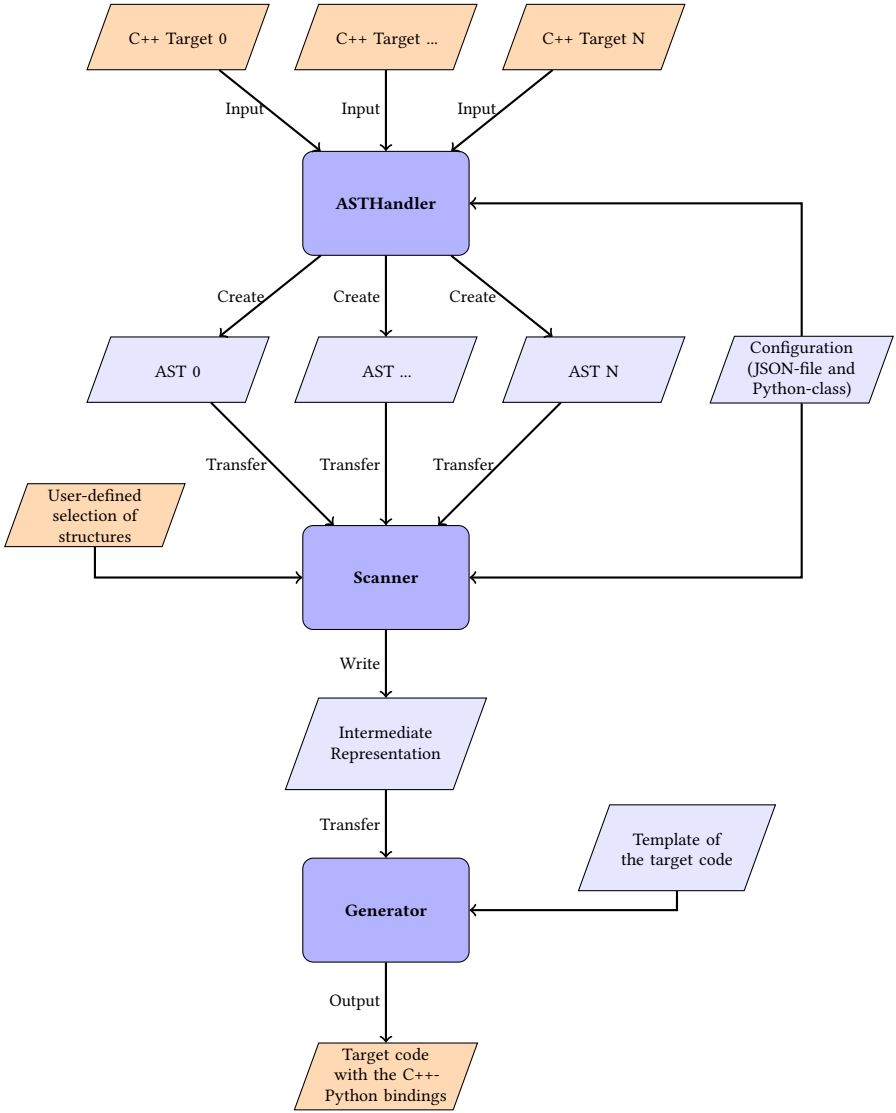

To systematically transform the recurring patterns, into reusable Python interfaces, the MEmilio interface generation approach has been developed as a multi-stage processing pipeline illustrated~\cref{fig:pygen_workflow_new}. The workflow consists of three central phases: structural extraction, intermediate representation, and template-based code generation. As the framework itself, the pipeline has been designed in a modular fashion such that the individual components can also serve multiple use cases.

In large and modular software libraries, it is uncommon to provide a single build target that covers the entire code base. Instead, functionality is typically distributed across several targets, each requiring only the components of a specific task. Leveraging this structure in MEmilio, we avoid the creation and maintenance of an additional target that includes all source files and which would have to be adapted with every new model and updated whenever the project structure changes. The processing pipeline therefore starts with multiple C++ targets, each representing a defined build target within the software backend and containing the corresponding components such as a model and its implementation files. The pipeline processes the existing targets individually and combines the extracted information to obtain a complete representation of the library. As shown on the right of~\cref{fig:pygen_workflow_new}, a centralized configuration component acts as the main interface for providing relevant paths and parameters to the processing stages of the \textit{ASTHandler} and the \textit{Scanner}. Based on this configuration, the ASTHandler locates the source code belonging to each C++ target and converts it into an AST representation using the appropriate compiler settings.

The ASTHandler serves as the central coordination unit of the first stage and controls the creation of the ASTs ensuring that each target is translated into an independent AST. For increasing numbers of input targets, the ASTHandler allows processing of the individual targets in parallel using Python's \texttt{multiprocessing} functionality. After generation, the ASTs are stored as structured representations and made available for subsequent analysis. The Scanner processes the ASTs sequentially and extracts all user-defined structures, such as selected classes and functions which are intended for the Python bindings. The extracted information is stored in a consolidated \textit{Intermediate Representation}, which serves as a stable interface between analysis and code generation. The Intermediate Representation serves as a flexible interface that shall allow the Scanner or \textit{Generator} to be exchanged independently. In the final step, the Generator consumes the information from the Intermediate Representation and applies predefined code templates to produce syntactically correct and consistent C++ binding code, which is written to dedicated output files.

Following the workflow illustrated in~\cref{fig:pygen_workflow_new}, the overall architecture of the code generation process emphasizes modularity and scalability. Each processing stage is clearly separated and communicates through well-defined data structures, which simplifies maintenance and allows individual components to be extended or replaced independently. In particular the strict separation between AST construction, analysis, and code generation ensures that changes in one phase do not propagate unintended side effects to subsequent stages.

\subsection{Extracting structural information}

\subsubsection{Transforming source code and organizing ASTs}
The transformation of source code into ASTs is preceded by a dedicated configuration phase that defines the structural context required for the analysis. This configuration specifies the source environment, including source files, output directories, and compiler-related settings required for AST construction. It ensures that the analyzed code is processed within the correct build context and that all dependencies can be resolved consistently.

The resulting AST provides a hierarchical representation of nodes of the syntactic structure of the source code, preserving structural and semantic relationships while abstracting from irrelevant formatting details. The ASTs form the basis for all subsequent analysis steps. Each node of the AST corresponds to a language construct such as a class, function, parameter, or expression, thereby enabling the systematic identification of recurring structures such as model parameter sets, flow specifications and model definitions. 

To manage the creation and organization of ASTs in our framework, we utilize the dedicated ASTHandler class. First, this class provides a unified interface for generating ASTs from source code. Second, the class maintains an internal data structure to manage the created ASTs. This includes functionality for adding new ASTs, removing obsolete ones, and accessing the stored ASTs for subsequent analysis or transformations. By centralizing both the creation and management of ASTs, the ASTHandler simplifies interaction with the underlying \textit{Clang} infrastructure and provides a scalable solution for analyzing large code bases. 

A central conceptual element of the AST approach is the introduction of unique identifiers and line numbers, enabling each node in the AST to be mapped precisely to a corresponding line in the source code and retrieved through an unambiguous identifier. This assignment of IDs and line numbers provides the structural basis for subsequent mechanisms. The design introduces the capability to selectively extract subtrees based on these identifiers. This allows specific code segments, such as individual classes or functions, to be isolated and visualized independently. Such selective visualization supports a more focused examination of individual program components by suppressing irrelevant structures and emphasizing the relevant context. This feature proves particularly valuable when investigating deeply nested constructs. The concept supports a flexible range of output formats. First, it supports a structured tree view within a text file or as a compact excerpt displayed in the console. Second and to facilitate a more effective analysis of ASTs, a dedicated visualization class was introduced. This component serves as the foundation for a set of mechanisms aimed at improving the interpretability and flexibility of AST representations.

In the following~\cref{fig:ASTGraphParameter,fig:simulate_ast}, we provide two examples of such AST representations for structures from the MEmilio framework, illustrating how a C++ function is transformed into its hierarchical AST structure and highlighting the structural decomposition of source code into analyzable language constructs. Together, these two representations demonstrate how the structural components of a function are stored.

\begin{example}[AST of \texttt{Parameters} constructor]
The top node of~\cref{fig:ASTGraphParameter} represents the declaration of the \texttt{Parameters} constructor as shown in Code Snippet~\ref{code:parameterlist} and from which several substructures branch out. These structures can be found in the AST in an hierarchical pattern where all nodes represent different \texttt{CursorKind}s that describe the kind of entity the cursor points to. Most prominently from the left, we see the \texttt{CursorKind.PARM\_DECL} which points to the declaration of the input parameter \texttt{num\_agegroups} as depicted in Code Snippet~\ref{code:parameterlist}. Below the \texttt{PARM\_DECL} cursor kind, we see the \texttt{CursorKind.TYPE\_REF} to the \texttt{mio::AgeGroup}, the type of the expected parameter. From the left, the second and third child of the top node encode the base class initialization, with a \texttt{TEMPLATE\_REF} to the base class \texttt{ParametersBase} and a \texttt{TYPE\_REF} the its corresponding template type parameter \texttt{FP}. The fourth child refers back to the first node, which is passed as an input parameter to the constructor of the base class. Similary, the next two child nodes of the constructor \texttt{MEMBER\_REF} and \texttt{CALL\_EXPR} represent the member initialization of \texttt{m\_num\_groups} through a call to the constructor of the class \texttt{AgeGroup}. The last child node \texttt{COMPOUND\_STMT} represents the body of the constructor as it can be seen in the constructor code itself. The compound statement \texttt{COMPOUND\_STMT} of the AST does not contain any further information, as its constructor body is empty.
\end{example}

\begin{figure}[!t]
  \centering
  \begin{tikzpicture}[scale=0.5,
    level distance=3cm,
    level 1/.style={sibling distance=3.5cm},
    level 2/.style={sibling distance=5cm},
    level 3/.style={sibling distance=5cm},
    edge from parent/.style={->,>=stealth,draw,thick},
    every node/.style={scale=0.45, font=\Large,
      ellipse, draw, minimum width=3cm, minimum height=1cm,
      align=center,thick
    }
  ]

  \node[fill=white, align=center] {CONSTRUCTOR \\ (Parameters<FP>)}
    child { node[fill=white] {PARM\_DECL \\ (num\_agegroups)}
        child { node[fill=white] {TYPE\_REF \\ (struct\\ mio::AgeGroup)}
    }}
    child { node[fill=white] {TEMPLATE\_REF \\ (ParametersBase)}
    } 
    child { node[fill=white] {TYPE\_REF \\ (FP)}
    }
    child { node[fill=white] {UNEXPOSED\\\_EXPR }
        child { node[fill=white] {DECL\_REF\\\_EXPR \\ (num\_agegroups)}
    }}
    child { node[fill=white] {MEMBER\_REF \\ (m\_num\_groups)}  
    }
    child { node[fill=white] {CALL\_EXPR \\ (AgeGroup)}
        child { node[fill=white] {UNEXPOSED\\\_EXPR\\ (num\_agegroups)}
            child { node[fill=white] {DECL\_REF\_EXPR \\ (num\_agegroups)}
    }}}
    child { node[fill=white] {COMPOUND\_STMT}};

  \end{tikzpicture}
  \caption{\textbf{Graph visualization of the \texttt{Parameters<FP>} constructor of the MEmilio framework transformed into AST.}}
  \label{fig:ASTGraphParameter}
\end{figure}

\begin{figure}[!b]
\centering
\begin{tikzpicture}[scale=0.5,
    level distance=2cm,
    level 1/.style={sibling distance=4cm},
    level 2/.style={sibling distance=3cm},
    level 3/.style={sibling distance=4cm},
    level 4/.style={sibling distance=3cm},
    edge from parent/.style={->,>=stealth,draw,thick},
    every node/.style={
        scale=0.45, font=\Large,
        ellipse,
        draw,
        thick,
        fill=white,
        align=center,
        minimum width=3.2cm,
        minimum height=1.2cm
    }
]
\node {FUNCTION\_TEMPLATE\\(simulate)}
    child[xshift=-0.25cm, edge from parent/.style={->, draw, dashed, thick}] {
        node[fill=white, style=dashed] {TEMPLATE\_TYPE\\\_PARAMETER\\{[...]}}
        child[edge from parent/.style={->, draw, dashed, thick}] { node[fill=white, style=dashed] {[...]} }                
    } 
    child[xshift=0.25cm] {
        node {TEMPLATE\_REF\\(TimeSeries)}
    } 
    child[xshift=0.25cm] {
        node {TYPE\_REF\\(FP)}
    } 
    child[edge from parent/.style={->, draw, dashed, thick}] {
        node[fill=white, style=dashed] {PARM\_DECL\\{[...]}}
        child[edge from parent/.style={->, draw, dashed, thick}] { node[fill=white, style=dashed] {[...]} }                
    }
    child[xshift=1cm]  { node[fill=white] {COMPOUND\_STMT} 
        child[yshift=-2cm] {
            node {CALL\_EXPR}
            child[edge from parent/.style={->, draw, dashed, thick}] { node[fill=white, style=dashed] {[...]} }
        }
        child[yshift=-2cm] {
            node {DECL\_STMT}
            child[edge from parent/.style={->, draw, dashed, thick}] { node[fill=white, style=dashed] {[...]} }
        }
        child[yshift=-2cm] {
            node {IF\_STMT}
            child[edge from parent/.style={->, draw, dashed, thick}] { node[fill=white, style=dashed] {[...]} }
        }
        child[yshift=-2cm] {
            node {CALL\_EXPR }
            child[edge from parent/.style={->, draw, dashed, thick}] { node[fill=white, style=dashed] {[...]} }
        }
        child[yshift=-2cm] {
            node {RETURN\_STMT}
            child[edge from parent/.style={->, draw, dashed, thick}] { node[fill=white, style=dashed] {[...]} }
        }
    };
\end{tikzpicture}
\caption{\textbf{Reduced graph visualization of the \texttt{simulate} function of the MEmilio framework transformed into AST.}}
\label{fig:simulate_ast}
\end{figure}

\begin{example}[Reduced AST of \texttt{simulate} function]
In~\cref{fig:simulate_ast}, we show a reduced graphical representation of the AST of the \texttt{simulate} function which has in total 54 nodes. The first node \texttt{TEMPLATE\_TYPE\_\linebreak
PARAMETER} is shown with dashed lines as it aggregates the declaration of the three template type parameters for a simplified visualization. The types of the actual nodes are replaced by concrete types during template instantiation. From the left, the next two nodes represent the function return type with \texttt{TEMPLATE\_REF} describing the template class and \texttt{TYPE\_REF} the corresponding template parameter as a reference to the template type of the simulate function. Finally, the fourth visualized node aggregates five actual nodes of~\texttt{PARM\_DECL}. Code Snippets~\ref{code:simulate}~and~\ref{code:ASTSubtreeSimulationOSEIR} show the corresponding parameter declarations of \textit{t0}, \textit{tmax}, \textit{dt}, \textit{model}, and \textit{integrator\_core}, where the latter obtains a standard argument. We also show the \texttt{COMPOUND\_STMT}, that contains several statements, namely a \texttt{CALL\_EXPR}, a \texttt{DECL\_STMT}, an \texttt{IF\_STMT}, and a \texttt{RETURN\_STMT}. The first \texttt{CALL\_EXPR} represents a call to a member function of the model. The \texttt{DECL\_STMT} declares the simulation object and initializes it using the model, \textit{t0}, and \textit{dt} parameters. The \texttt{IF\_STMT} checks whether an \texttt{integrator\_core} has been provided and, if so, assigns it to the simulation. The subsequent \texttt{CALL\_EXPR} advances the simulation up to \textit{tmax}. Finally, the \texttt{RETURN\_STMT} returns the resulting simulation data through a \texttt{CALL\_EXPR} corresponding to \texttt{mio::simulate}.
\end{example}

In Code Snippet~\ref{code:ASTSubtreeSimulationOSEIR}, we additionally provide the list-based subtree output. The \texttt{simulate} function has node identifier 754637. The list-based subtree includes the selected node as its root along with all of its direct and indirect descendants. 

\vspace{0.5cm}
\begin{minipage}[c]{0.96\textwidth}
\begin{lstlisting}[caption={\textbf{Reduced Subtree of an AST in textual representation.}},label={code:ASTSubtreeSimulationOSEIR}]
ID:754637 simulate <CursorKind.FUNCTION_TEMPLATE> ../compartments/simulation.h:100
    ID:754638 FP <CursorKind.TEMPLATE_TYPE_PARAMETER> ../compartments/simulation.h:99
    ID:754639 Model <CursorKind.TEMPLATE_TYPE_PARAMETER> ../compartments/simulation.h:99
    ID:754640 Sim <CursorKind.TEMPLATE_TYPE_PARAMETER> ../compartments/simulation.h:99
        ID:754641-754644 [...]
    ID:754645 TimeSeries <CursorKind.TEMPLATE_REF> ../compartments/simulation.h:100
    ID:754646 FP <CursorKind.TYPE_REF> ../compartments/simulation.h:100
    ID:754647 t0 <CursorKind.PARM_DECL> ../compartments/simulation.h:100
        ID:754648 FP <CursorKind.TYPE_REF> ../compartments/simulation.h:100
    ID:754649 tmax <CursorKind.PARM_DECL> ../compartments/simulation.h:100
        ID:754650 FP <CursorKind.TYPE_REF> ../compartments/simulation.h:100
    ID:754651 dt <CursorKind.PARM_DECL> ../compartments/simulation.h:100
        ID:754652 FP <CursorKind.TYPE_REF> ../compartments/simulation.h:100
    ID:754653 model <CursorKind.PARM_DECL> ../compartments/simulation.h:100
        ID:754654 Model <CursorKind.TYPE_REF> ../compartments/simulation.h:100
    ID:754655 integrator_core <CursorKind.PARM_DECL> ../compartments/simulation.h:101
        ID:754656-754660 [...]
    ID:754661 <CursorKind.COMPOUND_STMT> ../compartments/simulation.h:102
        ID:754662 <CursorKind.CALL_EXPR> ../compartments/simulation.h:103
            ID:754663 <CursorKind.MEMBER_REF_EXPR> ../compartments/simulation.h:103
                ID:754664 model <CursorKind.DECL_REF_EXPR> ../compartments/simulation.h:103
        ID:754665 <CursorKind.DECL_STMT> ../compartments/simulation.h:104
            ID:754666 sim <CursorKind.VAR_DECL> ../compartments/simulation.h:104
                ID:754667 Sim <CursorKind.TYPE_REF> ../compartments/simulation.h:104
                ID:754668 <CursorKind.UNEXPOSED_EXPR> ../compartments/simulation.h:104
                    ID:754669-754671 [...]
        ID:754672 <CursorKind.IF_STMT> ../compartments/simulation.h:105
            ID:754673 integrator_core <CursorKind.DECL_REF_EXPR> ../compartments/simulation.h:105
            ID:754674 <CursorKind.COMPOUND_STMT> ../compartments/simulation.h:105
                ID:754675-754682 [...]
        ID:754683 <CursorKind.CALL_EXPR> ../compartments/simulation.h:108
            ID:754684 <CursorKind.MEMBER_REF_EXPR> ../compartments/simulation.h:108
                ID:754685 sim <CursorKind.DECL_REF_EXPR> ../compartments/simulation.h:108
            ID:754686 tmax <CursorKind.DECL_REF_EXPR> ../compartments/simulation.h:108
        ID:754687 <CursorKind.RETURN_STMT> ../compartments/simulation.h:109
            ID:754688 <CursorKind.CALL_EXPR> ../compartments/simulation.h:109
                ID:754689 <CursorKind.MEMBER_REF_EXPR> ../compartments/simulation.h:109
                    ID:754690 sim <CursorKind.DECL_REF_EXPR> ../compartments/simulation.h:109
\end{lstlisting}
\end{minipage}

\subsubsection{Identifying relevant structures from ASTs}

The Scanner is the central component responsible for analyzing ASTs and extracting relevant structures from them. It systematically traverses the ASTs, identifies key elements according to predefined criteria and collects the information necessary for subsequent processing stages in a dictionary. By isolating and organizing these structures, the Scanner enables efficient and structured handling of the program's syntactic and semantic information. To determine which elements of the AST should be extracted, the Scanner is equipped with a configuration dictionary (see Code Snippet~\ref{code:configdict}) in which all target structures are defined in advance. This dictionary specifies types, function names, class identifiers, or syntactic constructs that are relevant for the analysis and later code generation stages. 

While high-level dictionary input referring to an entire model automatically identifies and extracts large sets of AST nodes, user-defined configuration also allows individual functions to be selected. In Code Snippet~\ref{code:configdict}, we, for instance, provide the short and simple notation for the simulate function to be added to the model components that are identified automatically. With this approach, we allow, on one hand, simplified and automatic selection of all model components while, on the other hand, we maximize flexibility for instances where automatic deduction from model level is not sufficient. Additionally defining the desired function in the configuration dictionary extends the set of extracted components according to the requirements of a particular analysis or code generation task.

\vspace*{0.5cm}
\begin{minipage}[c]{0.96\textwidth}
\begin{lstlisting}[caption={\textbf{User-defined selection for simulate function.}},label={code:configdict}]
    {
        "type": "function",
        "name": "simulate"
    }
\end{lstlisting}
\end{minipage}

The traversal of the AST is performed using a depth-first search (DFS) approach. Each node in the tree is visited, starting from the root node and exploring as far as possible along each branch before backtracking. This method ensures that all nodes and their hierarchical relationships are examined in a structural manner. The DFS traversal is implemented recursively, allowing the Scanner to efficiently process nested structures and extract the relevant information from every node within the AST. At each visited node the Scanner compares the nodes' attributes with the specifications contained in the configuration dictionary. If the node matches an entry defined in the dictionary the Scanner triggers the extraction step. The extracted information is first stored in a temporary dictionary called \textit{found\_info}. Each completed \textit{found\_info} dictionary is then appended to a list named \textit{found\_bindings} which accumulates all discovered and processed bindings throughout the traversal. This mechanism ensures that the Scanner produces a structured and easy accessible representation of all elements that match the predefined search criteria.

\subsection{Organization of information in Intermediate Representation}
After identifying and extracting the relevant information from the AST, the data must be stored in an intermediate form before it can be used by the Generator. To do so, we introduce our Intermediate Representation, which serves as the data model between the scanning and generation stages. Its purpose is to organize and preserve the extracted information in a structured manner, allowing the subsequent generation step to access and process it. With the Intermediate Representation, the Generator can operate exclusively on this structured model, no longer needing to access or traverse the original ASTs. 

\vspace*{0.5cm}
\begin{minipage}[c]{0.96\textwidth}
\begin{lstlisting}[caption={\textbf{Intermediate Representation entry.} },label={code:found_info}]
found_info = binding_type_info(
        type=binding_type,
        name=node.spelling,
        cursorkind=node.kind.name,
        namespace=namespace,
        return_type=node.result_type.spelling if node.kind.is_declaration() else "",
        arg_types=function_arguments.get("arg_types", []),
        arg_names=function_arguments.get("arg_names", []),
        parent_name=function_arguments.get("parent_name",""),
        is_const=function_arguments.get("is_const", False),
        is_member=function_arguments.get("is_member", False),
        methods=[]
)
\end{lstlisting}
\end{minipage}

The \textit{found\_info} dictionary in Code Snippet~\ref{code:found_info} illustrates the creation of one entry in the Intermediate Representation. For each relevant declaration encountered during the AST traversal, an object of type \textit{binding\_type\_info} is instantiated. The object stores essential metadata, including the declaration type, name, cursor kind, and namespace. For function declarations, additional information such as the return type, parameter types, and parameter names are recorded. Furthermore, properties describing the declaration context, for instance whether the function is a member or independent, are included. In another step, class declarations collect their associated member functions.

The Intermediate Representation in Code Snippet~\ref{code:IR} is tailored to the description of a (new) model implemented in MEmilio and it combines configuration information, model-specific properties, and the bindings extracted during the AST analysis. While attributes such as \textit{model\_class}, \textit{namespace}, and \textit{scalartype} describe the model and generation context, collections such as \textit{model\_init}, \textit{population\_groups} and \textit{include\_list} preserve structural information required during code generation. The \textit{found\_bindings} field provides the connection to the individual \textit{binding\_type\_info} entries. It stores all relevant declarations identified during the scanning phase and therefore forms the main input for the binding generation process.

\vspace{0.5cm}
\begin{minipage}[c]{0.96\textwidth}
\begin{lstlisting}[caption={\textbf{Structure of the Intermediate Representation.}},label={code:IR}]
@dataclass
class IntermediateRepresentation:
    namespace: str = ""
    model_class: str = ""
    python_module_name: str = ""
    parameterset: str = ""
    parameterset_wrapper: str = ""
    simulation: bool = False
    is_compartmentalmodel: bool = False
    is_flowmodel: bool = False
    has_age_group: bool = False
    has_draw_sample: bool = False
    scalartype: str = "double"
    python_generation_module_path: str = ""
    target_folder: str = ""
    enum_populations: dict = field(default_factory=dict)
    model_init: list = field(default_factory=list)
    model_base: list = field(default_factory=list)
    model_base_templates: str = ""
    population_groups: list = field(default_factory=list)
    include_list: list = field(default_factory=list)
    age_group: dict = field(default_factory=dict)
    found_bindings: list[binding_type_info] = field(
        default_factory=list)
    ...
\end{lstlisting}
\end{minipage}


\subsection{Interface generation}

In this section, we first present the model-specific binding generation in~\cref{sec:mode_specific_gen} as required by our initial use case. In~\cref{sec:general_cpp_gen}, we present a generalization of binding code generation, considering general C++ structures, to be used either independently or complementary to the model-specific generation..

\subsubsection{Model-specific generation}\label{sec:mode_specific_gen}

After the successful identification and extraction of relevant structures within the Intermediate Representation, the next step is the generation of bindings. The binding generation phase uses a template-based representation to transform the structural information collected by the Scanner into concrete binding code. The template shown in Code Snippet~\ref{code:placeholderfile} defines the overall structure of the generated \textit{PYBIND11\_MODULE} and contains placeholders for model-specific binding components. For each placeholder, of the form \texttt{\$\{name\_of\_the\_placeholder\}}, the Generator provides a function that receives the Intermediate Representation and returns the corresponding string. Some placeholders, for instance \texttt{\$\{namespace\}}, \texttt{\$\{parameterset\}}, and \texttt{\$\{ScalarType\}}, appear inside a binding call and are replaced by the corresponding C++ type name. Others, such as \texttt{\$\{model\_init\}}, \texttt{\$\{simulation\}}, and \texttt{\$\{draw\_sample\}}, stand on their own and are replaced by complete code segments. The placeholder \texttt{\$\{simulation\}} illustrates how the generated code adapts to the model-specific structures. For instance, if the model is defined as a FlowModel, the binding statement exposes the class and an additional function for simulating with flows. The function composing these binding statements used to replace the placeholder, including all possible variants, is shown in Code Snippet~\ref{code:placeholder_function} in the appendix. The template itself remains independent of a particular model, while the generated binding code is specialized to the structures identified for the respective model.  

\vspace{0.5cm}
\begin{minipage}[c]{0.96\textwidth}
\begin{lstlisting}[caption={\textbf{Placeholders for the binding text.}},label={code:placeholderfile}]
[...]
PYBIND11_MODULE(_simulation_${python_module_name}, m)
{
    ${population_enums}

    ${parameterset_indexing}
    pymio::bind_ParameterSet<${namespace}${parameterset}<${ScalarType}>, pymio::EnablePickling::Required>(m, "${parameterset}");

    ${parameterset_wrapper}
   
    pymio::bind_Population(m, "Populations", mio::Tag<${namespace}${model_class_name}<${ScalarType}>::${population}>{});
   
    pymio::bind_CompartmentalModel<${model_base_templates}>(m, "${model_class_name}Base");
    py::class_<${namespace}${model_class_name}<${ScalarType}>, pymio::EnablePickling::Required, ${model_base_templates}>(m, "${model_class_name}");
    
    ${model_init}

    ${simulation}
    
    ${draw_sample}
    [...]
}
\end{lstlisting}
\end{minipage}

\subsubsection{Generalization to structural C++ level}\label{sec:general_cpp_gen}

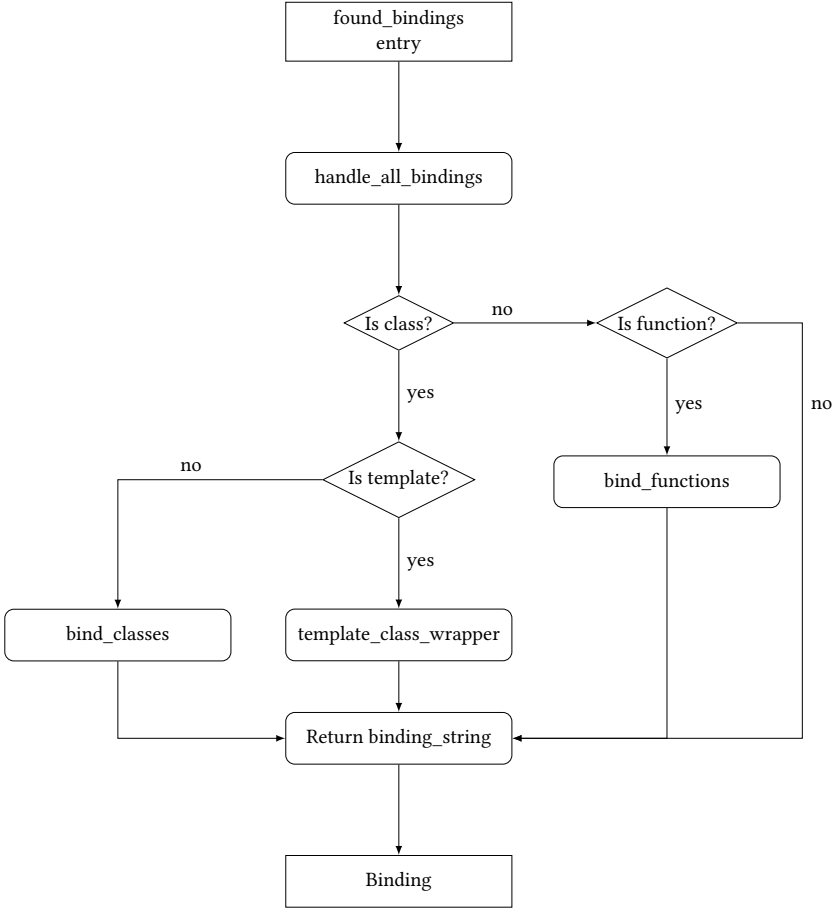
\begin{figure}[!h]
\centering
\resizebox{0.8\linewidth}{!}{%
\begin{tikzpicture}[
    scale=0.5,
    node distance=1.4cm,
    >=latex,
    every node/.style={font=\small},
    process/.style={
        draw,
        rounded corners,
        minimum width=3.5cm,
        minimum height=0.8cm,
        align=center
    },
    decision/.style={
        diamond,
        draw,
        aspect=2,
        align=center,
        inner sep=1pt
    },
    io/.style={
        draw,
        minimum width=3.5cm,
        minimum height=0.8cm,
        align=center
    }
]

\node[io] (input) {found\_bindings\\entry};

\node[process, below=of input] (handle) {handle\_all\_bindings};

\node[decision, below=of handle] (class) {Is class?};

\node[decision, below=of class] (template) {Is template?};

\node[process, below left=1.7cm and 2cm of template] (bindclass) {bind\_classes};

\node[process, below=of template] (templatewrap) {template\_class\_wrapper};

\node[decision, right=2.2cm of class] (function) {Is function?};

\node[process, below=1.5cm of function] (bindfunc) {bind\_functions};

\node[process, below=3cm of template] (return) {Return binding\_string};

\node[io, below=of return] (output) {Binding};

\draw[->] (input) -- (handle);
\draw[->] (handle) -- (class);

\draw[->] (class) -- node[right]{yes} (template);

\draw[->] (class.east) -- ++(1.5,0)
    |- node[pos=0.25,above]{no} (function);

\draw[->] (template) -- node[right]{yes} (templatewrap);

\draw[->] (template.west) -- ++(-1.8,0)
    -| node[pos=0.25,above]{no} (bindclass.north);

\draw[->] (function) -- node[right]{yes} (bindfunc);

\draw[->] (function.east) -- ++(2,0)
    |- node[pos=0.25,above,xshift=3mm,yshift=1.75cm]{no} (return);

\draw[->] (bindclass) |- (return);
\draw[->] (templatewrap) -- (return);
\draw[->] (bindfunc) |- (return);

\draw[->] (return) -- (output);

\end{tikzpicture}
}
\caption{\textbf{Template selection and generation workflow.}}
\label{fig:template-selection-workflow}
\end{figure}

Either independently or combined with the model-specific generation, we can generate selected parts of the code with the user-defined selection of structures described in~\cref{sec:concept_and_workflow}. This more versatile approach for semi-automated code generation of prominent C++ structures such as classes, templated classes, or functions enables more flexibility but also requires more user-specific input. As in the previous section, the generated code is based on templates. However, we no longer work with a fixed template structure but require the selection of a suitable template for each extracted element. 

To select the appropriate template for each extracted structure, the system employs a coordination component as visualized in~\cref{fig:template-selection-workflow}. This component iterates over all entries identified and integrated in the Intermediate Representation and determines which template category should be applied. The decision is based primarily on the general type of the extracted structure and when necessary on more fine grained classification attributes.

More precisely, class like structures are first distinguished according to their specific form, enabling the system to separate classes from functions. Each category is then forwarded to a specialized processing routine that is responsible for generating the corresponding binding representation. Function like structures undergo a similar process and are routed to dedicated handling mechanisms that evaluate their characteristics and choose an appropriate generation strategy.

This separation of responsibilities ensures that the extracted information is placed into the correct semantic context and establishes a clear link between the analysis phase and the code generation phase. It allows the simple expansion to previously nonconsidered structures, also including novel functionality of C++ or the binding library. The template selection mechanism acts as the core decision layer that governs how the abstract information is transformed into concrete bindings.

Once an element has been assigned to a template, a dedicated generation procedure processes the relevant properties associated with that element. These properties such as names, signatures, or structural relationships form the foundation for determining how the resulting binding code should be composed. During this step the system evaluates specific conditions to identify the correct generation strategy. Structures with special characteristics such as overloaded functions or otherwise not directly exposable functions are handled using more flexible or adaptive template variants. Depending on the structural context the generated representation may vary, e.g., between methods that belong to a class and standalone functions as shown in Code Snippet~\ref{code:template-function}, ensuring that each binding is correctly integrated into its surrounding module or object hierarchy. Note that standalone as well as member functions always obtain a \texttt{scalartype} template which is defined by the utilized computation precision in MEmilio.

\vspace{0.5cm}
\begin{minipage}[c]{0.96\textwidth}
\begin{lstlisting}[caption={\textbf{Template used for generating bindings of standalone and member functions.}},label={code:template-function}]
def direct_binding(name: str, namespace: str, scalartype: str, py_args: str, prefix: str, is_member: bool, parent_name: str) -> str:
    
    if not is_member:
        return (
            f'\t{prefix}.def('
            f'\n\t\t"{name}",'
            f'\n\t\t&{namespace}{name}<{scalartype}>,'
            f'\n\t\t{py_args}\n\t);\n\n'
        )
    else:
        return (
            f'\t{prefix}.def("{name}", &{namespace}{parent_name}<{scalartype}>::{name})\n'
        )
\end{lstlisting}
\end{minipage}

The combination of AST analysis, the Intermediate Representation, and general structure-based code generation forms a complete transformation pipeline. Once all preceding processing steps have been executed, the resulting output is the final binding code, an example of which is shown in Code Snippet~\ref{code:template-simulate-function}.

\vspace{0.5cm}
\begin{minipage}[c]{0.96\textwidth}
\begin{lstlisting}[caption={\textbf{Generated binding for the simulate function.}},label={code:template-simulate-function}]
m.def(
		"simulate",
		&mio::simulate<double>,
		py::arg("t0"), py::arg("tmax"), py::arg("dt"), py::arg("model"), py::arg("integrator_core")
	);
    \end{lstlisting}
\end{minipage}

\section{Results}
\label{sec:results}

In the previous sections, we have presented our approach for providing a Python frontend to the efficient C++ backend in MEmilio. For users, the corresponding Python shall deliver a flexible frontend that does not stand back the C++ performance when executing MEmilio's models and that allows to couple established Python frameworks for, e.g., sensitivity analysis or parameter estimation. For developers, the semi-automated generation approach shall deliver a fast and flexible way of binding functional extensions or new models. We demonstrate that our approach is indeed capable of addressing these requirements in individual sections.

\subsection{Runtime overhead of the Python bindings}
\label{sec:binding-performance}

We first show that the overhead of the Python bindings, despite its flexibility in setting up user-specific simulations, is negligible. Therefore, we consider a graph-based epidemic simulation where, at every graph node, the local infection dynamics are represented by an age-stratified ODE-based SECIR-type model. Individuals progress from susceptible through exposed but not yet infectious, infectious without symptoms, and symptomatic infectious states. The model additionally resolves severe and critical infections, corresponding to hospitalization and intensive-care treatment, as well as recovery and death. The local models are coupled through the graph edges, which describe commuter mobility occurring at discrete time points twice a day. Local infection dynamics are advanced between these discrete mobility events. The parameters are taken from~\cite{bicker2026memilio} and further details on the model are given in~\cite{ZUNKER2027_stage_aligned}.

We consider graphs with 10 to 1000 nodes where 40~\% of the edges represent nontrivial mobility exchange, all other edges are removed. The local models are all initialized with either $G\in\{1,3,6\}$ age groups. All simulations advance the system from $t=0$ to $t=1$ with a fourth-order Runge-Kutta method of step size $\Delta t$. For each configuration, 30 measurements were performed after one warm-up in each of three independent jobs. The reported values are medians of the per-job medians, and the error bars span their minimum and maximum. We measure only the time for the simulate itself as the initialization is excluded from timing.  All measurements use one CPU core on an exclusively allocated node with two AMD EPYC 9555 64-Core processors.

\begin{figure}[!ht]
    \centering
    \includegraphics[width=1.\linewidth]{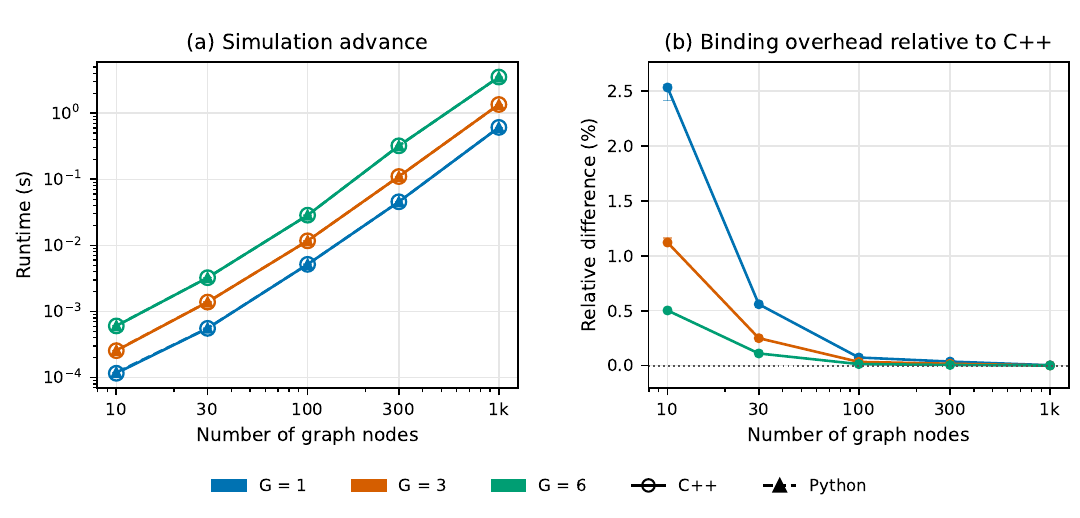}
    \caption{\textbf{Runtime of the graph-based ODE-SECIR simulation in Python and C++ (left) and relative Python-binding overhead (right) for a step size of $\Delta t=0.1$.} Panel~(a) compares the C++ runtime with the runtime obtained for execution for Python. Panel~(b) shows their relative difference. $G$ denotes the number of age groups.}
    \label{fig:graph-bindings-dt01}
\end{figure}

\begin{figure}[!ht]
    \centering
    \includegraphics[width=1.\linewidth]{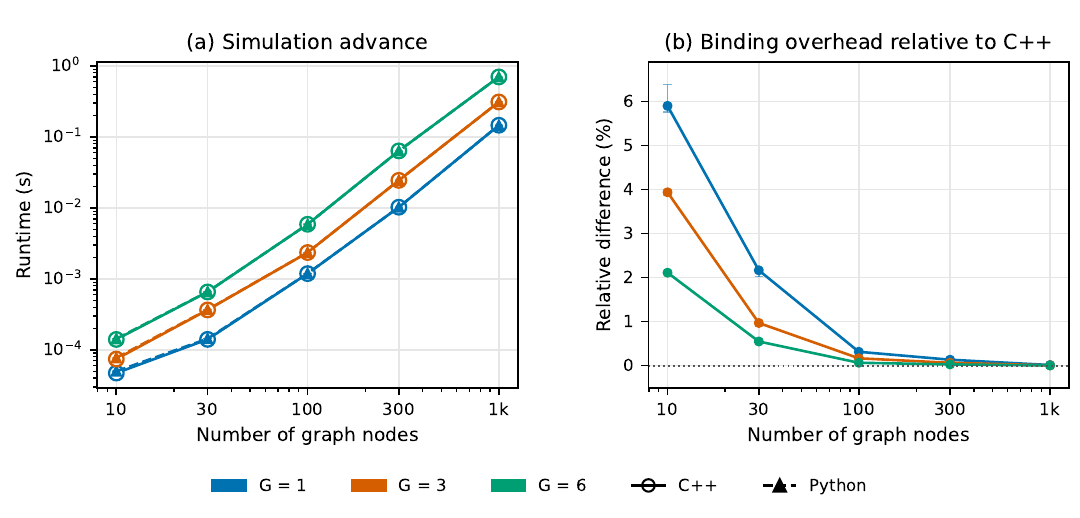}
    \caption{\textbf{Runtime of the graph-based ODE-SECIR simulation in Python and C++ (left) and relative Python-binding overhead (right) for a step size of $\Delta t=0.5$.} Panel~(a) compares the C++ runtime with the runtime obtained for execution for Python. Panel~(b) shows their relative difference. $G$ denotes the number of age groups.}
    \label{fig:graph-bindings-dt05}
\end{figure}

\Cref{fig:graph-bindings-dt01,fig:graph-bindings-dt05} compare the runtimes for simulating the same scenarios directly in C++ against running the simulation through Python bindings for different step sizes. Panel~(a) shows that the runtime increases with both the number of graph nodes and the number of age groups. The C++ and Python markers nearly coincide because both execute the same C++ backend simulation code. Panel~(b) isolates the cost of the Python execution and demonstrate the comparatively small cost of across all simulations.

For simulations with 10 graph nodes, the absolute overhead only ranges around $3\,\mu\mathrm{s}$, which is too small to be visible in Panel~(a). With $\Delta t=0.1$, this corresponds to an overhead between $0.5~\%$ and $2.5~\%$, depending on the number of age groups. For $\Delta t=0.5$, fewer Runge-Kutta steps are performed within the same bound function and the relative overhead at 10 nodes ranges from approximately $2.1~\%$ to $5.9~\%$. 

The differences becomes even less relevant as the graph size increases. At 30 ($\Delta t=0.1$) or 100 ($\Delta t=0.5$) nodes, the relative overhad is already below 1~\%. At 1000 nodes, the absolute overhead is approximately $23$--$28\,\mu\mathrm{s}$ for $\Delta t=0.1$ and $20$--$23\,\mu\mathrm{s}$ for $\Delta t=0.5$. Relative to the simulation runtime, this corresponds to only $0.0008~\%$--$0.0038~\%$ for $\Delta t=0.1$ and $0.0033~\%$--$0.0137~\%$ for $\Delta t=0.5$. The larger relative values for $\Delta t=0.5$ result from the smaller amount of computational work performed per Python call rather than from a larger absolute binding cost. Overall, invoking the graph simulation through Python introduces no practically relevant overhead for computationally substantial configurations. Moreover, the C++ and Python results were bitwise identical in all validation runs.

\subsection{Coupling with established Python frameworks}

\begin{figure} 
\begin{center}
\begin{minipage}{0.45\textwidth}
\includegraphics[width=\textwidth]{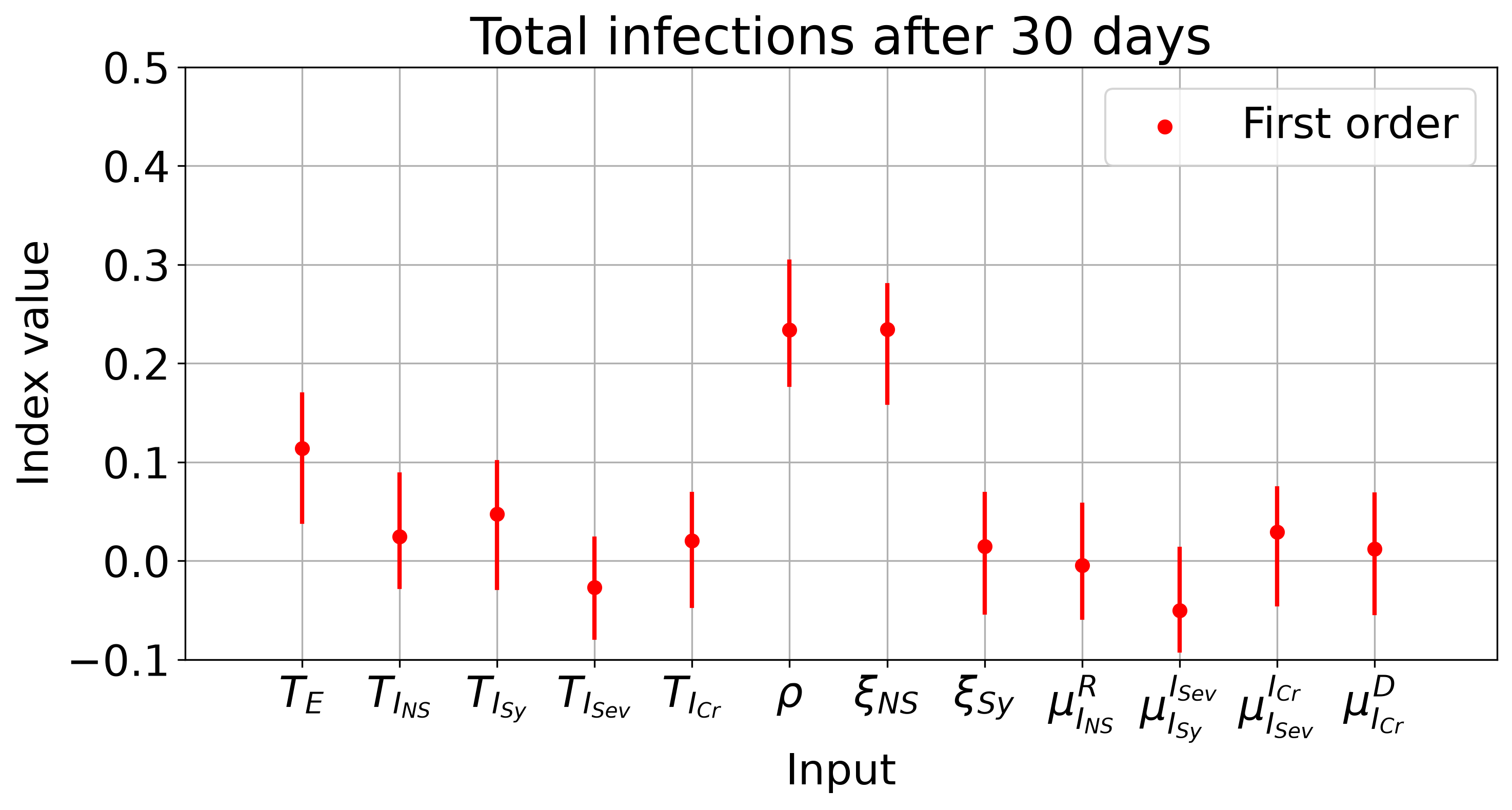}
\end{minipage}
\hspace*{0.5cm}
\begin{minipage}{0.45\textwidth}
\includegraphics[width=\textwidth]{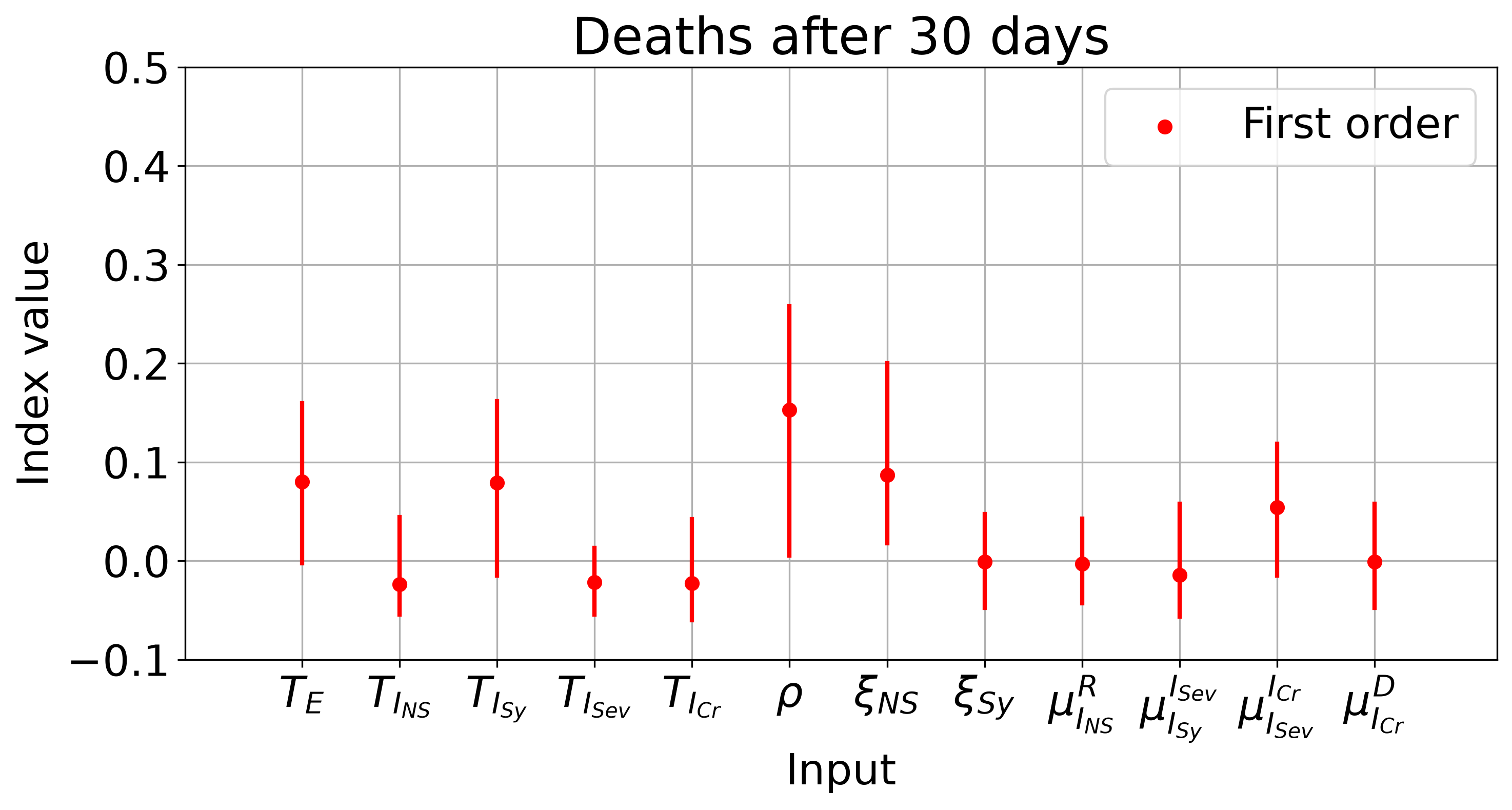}
\end{minipage}
	\caption{\textbf{Sobol' indices for the ODE-SECIR model with respect to the total number of infections (left) and deaths (right) after 30 days of simulation.}
    }
	\label{fig:sobol}
\end{center}
\end{figure}

Aside from providing a good performance, our Python bindings also enable the coupling of MEmilio with established Python frameworks. While selected MEmilio models have already been coupled successfully with advanced parameter estimation pipelines using BayesFlow~\cite{bayesflow_2023_software} or pyABC~\cite{schalte_pyabc_2022} (see~\cite{bicker2026memilio,Bicker_coupled_2025}), we here demonstrate how a (novel) MEmilio model can be swiftly connected to, e.g., a sensitivity analysis pipeline, only requiring a couple of additional lines. OpenTURNS~\cite{Baudin2016} is an open-source initiative for the treatment of uncertainties and allows to compute first-order Sobol' indices to assess how the variation in input parameter $x$ alone explains the output variance. We consider an ODE-based SECIR-type model as in the previous section and compute Sobol' sensitivity indices using a rank-based algorithm. We take the weighted average values as given in~\cite[Table~2]{schmidt_gnn_2025} and assume uniform marginal distributions with lower bound at 50~\% of the average value and upper bound at 150~\% of the average. The corresponding parameters and their values are also reprinted in \cref{tab:COVID-19parameters}. We consider a joint distribution made up by the 12 independent parameter distributions. Computing 1000 samples, we estimate Sobol' indices with respect to the total number of infections and deaths after 30 days of simulation. The code is provided with MEmilio\footnote{\url{https://github.com/SciCompMod/memilio/blob/pybind_paper_example_openturns_sobol/pycode/examples/simulation/ode_secir_simple_sensitivity.py}}. While the focus here is not on the assessment of the model, we briefly comment the results in~\cref{fig:sobol}. The transmission probability on contact $\rho$ is observed as the most influential parameter, while the analysis also correctly assigns a high importance to the transmission by nonsymptomatic individuals ($\xi_{I_{NS}}$). Also, we see that the shares (or deterministic probabilities) of severe and critical cases are observed to have little to no influence on the total number of infections while these parameters are as expected more important for the total number of deaths.

\subsection{Generating bindings to a new model}

\definecolor{okabeBlue}{HTML}{0072B2}
\definecolor{okabeOrange}{HTML}{D55E00}
\definecolor{okabeGreen}{HTML}{009E73}
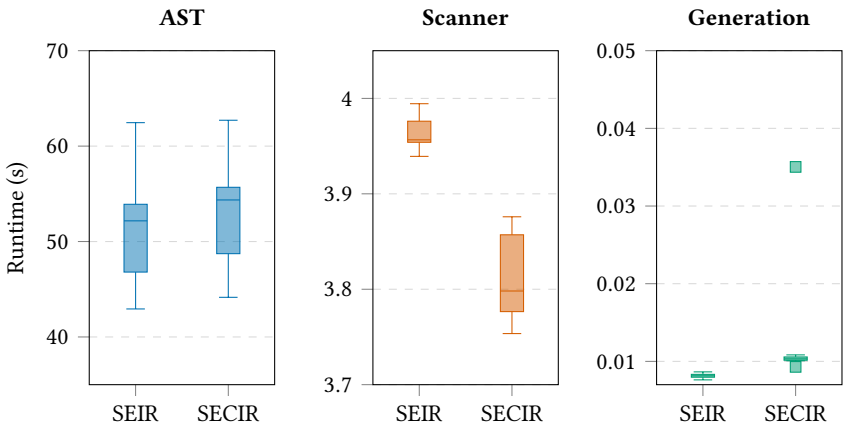
\begin{figure}[!b]
    \centering
    \begin{tikzpicture}
        \begin{groupplot}[
            group style={
                group size=3 by 1,
                horizontal sep=1.3cm,
            },
            boxplot/draw direction=y,
            boxplot/box extend=0.25,
            width=0.29\textwidth,
            height=6cm,
            xtick={1,2},
            xticklabels={SEIR, SECIR},
            x tick label style={font=\small},
            ymajorgrids=true,
            grid style={dashed, gray!30},
            tick align=outside,
            title style={font=\small\bfseries},
            xtick pos=bottom,
            ytick pos=left,
            ylabel style={font=\small},
            tick label style={font=\small},
            scaled y ticks=false,
            yticklabel style={
                 /pgf/number format/fixed,
                 /pgf/number format/precision=2,
            },
        ]

        \nextgroupplot[
            title={AST},
            ylabel={Runtime (s)},
            ymin=35,
            ymax=70,
            xmin=0.5,
            xmax=2.5,
        ]

        \addplot+[
            boxplot={draw position=1},
            fill=okabeBlue,
            fill opacity=0.5,
            draw=okabeBlue,
        ] table[row sep=\\, y index=0] {
            data\\
            52.409061\\
            51.854615\\
            53.130321\\
            50.485698\\
            52.168630\\
            54.676813\\
            43.114134\\
            57.884308\\
            42.935860\\
            62.458368\\
        };

        \addplot+[
            boxplot={draw position=2},
            fill=okabeBlue,
            fill opacity=0.5,
            draw=okabeBlue,
        ] table[row sep=\\, y index=0] {
            data\\
            52.709730\\
            52.204342\\
            54.856758\\
            54.357812\\
            54.627584\\
            56.510649\\
            45.253623\\
            57.504993\\
            44.153992\\
            62.710404\\
        };

        \nextgroupplot[
            title={Scanner},
            ymin=3.70,
            ymax=4.05,
            xmin=0.5,
            xmax=2.5,
        ]

        \addplot+[
            boxplot={draw position=1},
            fill=okabeOrange,
            fill opacity=0.5,
            draw=okabeOrange,
        ] table[row sep=\\, y index=0] {
            data\\
            3.989700\\
            3.994442\\
            3.939243\\
            3.962491\\
            3.957194\\
            3.955414\\
            3.993705\\
            3.956016\\
            3.956582\\
            3.952610\\
        };

        \addplot+[
            boxplot={draw position=2},
            fill=okabeOrange,
            fill opacity=0.5,
            draw=okabeOrange,
        ] table[row sep=\\, y index=0] {
            data\\
            3.807999\\
            3.753550\\
            3.777045\\
            3.853724\\
            3.866765\\
            3.798166\\
            3.793465\\
            3.776090\\
            3.860394\\
            3.875940\\
        };

        \nextgroupplot[
            title={Generation},
            ymin=0.007,
            ymax=0.05,
            xmin=0.5,
            xmax=2.5    
        ]

\addplot+[
    boxplot={draw position=1},
    fill=okabeGreen,
    fill opacity=0.5,
    draw=okabeGreen,
    mark options={draw=okabeGreen, fill=okabeGreen},
] table[row sep=\\, y index=0] {
    data\\
    0.008648\\
    0.007618\\
    0.008438\\
    0.007965\\
    0.007958\\
    0.008141\\
    0.008007\\
    0.008560\\
    0.008150\\
    0.008218\\
};

\addplot+[
    boxplot={draw position=2},
    fill=okabeGreen,
    fill opacity=0.5,
    draw=okabeGreen,
    mark options={draw=okabeGreen, fill=okabeGreen},
] table[row sep=\\, y index=0] {
    data\\
    0.035036\\
    0.010642\\
    0.010081\\
    0.010245\\
    0.010452\\
    0.010213\\
    0.010325\\
    0.009291\\
    0.010496\\
    0.010835\\
};

        \end{groupplot}
    \end{tikzpicture}

    \caption{\textbf{Runtime distributions of the entire workflow for SEIR and SECIR models.} Boxplots show the median, interquartile range and whiskers for AST creation (blue), scanning process (orange) and code generation (green). The panels use different y-axis scales to improve readability.}
    \label{fig:runtimes}
\end{figure}

As part of our results, we exemplarily show how the bindings of a (new) model can be generated with the \texttt{memilio-generation} package. For the Bindings Generator, we use the Clang C++ library together with the \texttt{libclang} Python bindings in required version 18.1.1, the Python packages \texttt{scikit-build}, \texttt{dataclasses}, \texttt{dataclasses\_json}, \texttt{graphviz} and \texttt{importlib-resources}, as well as the Python development libraries (\texttt{python3.x-dev}). The package installation with its dependencies is installed from the MEmilio repository\footnote{\url{https://github.com/SciCompMod/memilio}}, directory \texttt{pycode/memilio-generation}. Installation instructions are also provided in the documentation of the MEmilio Generation package with the Bindings Generator\footnote{\url{https://memilio.readthedocs.io/en/latest/python/m-generation.html}}.

We configure the generator for the SEIR model using the example script \texttt{example\_oseir.py}\footnote{\url{https://github.com/SciCompMod/memilio/blob/main/pycode/memilio-generation/memilio/tools/example_oseir.py}} for a single C++ target and AST. For the SECIR-type target, we only adapt source file path.

For comparison we consider the generation of an ODE-based SEIR and SECIR-type model, with both having handwritten bindings we can validate to. The runtime measurements of the entire process are split into building the AST of the C++ target, the Scanner extracting all relevant model information and lastly the generation of the binding code by the Generator. The benchmark is executed on a nonexclusive, single core of an Ubuntu 24.04 based work station with two Intel(R) Xeon(R) Gold 6230R CPU @ 2.10GHz (26 Cores, 52 Threads) CPUs. We conduct ten runs per model to obtain a more robust timing.

Figure~\ref{fig:runtimes} provides the results and shows that in both cases the runtime of the entire workflow takes on average less than one minute on a single core. More precisely, we obtain an average of 52\,s for the SEIR model and 53\,s for the SECIR model. Even in the slowest measured run, where system components might have interacted, the generation took less than 70 seconds. The largest part of this time is spent on the creation of the AST, while the Scanner itself requires only about four seconds, and the time needed for the generation itself is negligible.

\section{Conclusion}
\label{sec:conclusion}

In this paper, we have presented an approach for the (semi-automatical) provision of user-friendly interfaces for efficient simulation backends. Based upon pybind11, we bind the C++ high performance Modular EpideMIcs simuLatIOn software MEmilio in a Python interface, translating MEmilio's functionalities in a Pythonic way. Instead of high-level coupling of single executables, we strive for a maximum of flexibility. Providing the framework's major functionalities in Python, users can set up scenarios, choose model stratifications, manipulate model parameters, simulate ensemble settings, and couple established Python frameworks for sensitivity analysis or parameter estimation. 

Binding all relevant functionality, we demonstrate numerically that our holistic coupling approach represents minimal to no overhead and yields almost the same performance as our C++ baseline. Addressing Python developers as the largest group of developers and users in mathematical-epidemiological modeling, we effectively reduce the barriers for simulation-based pandemic mitigation. The latter holds in particular true as MEmilio's modular resolution in demography and space enables advanced, yet swift epidemic simulations from Python.

Through our suggested three stage process of semi-automatically generating Python bindings, we allow for the direct utilization of extensions and generalizations of the backend in Python. Translating a set of relevant C++ executables into abstract syntax trees (ASTs), we provide the base for a Scanner that flexibly extracts information corresponding to a user-defined configuration. Based upon an Intermediate Representation, a Generator generates new binding code based on a templated target code. While initially build for binding entirely new models, the Generator has been extended to create more general code based on structural C++ level. By replacing repetitive implementation work through automation and decoupling interface creation from the individual extension of the backend, our semi-automated approach establishes a scalable mechanism for extending the Python interface of MEmilio. Importantly, the entire process requires less than one minute of a single core of a Linux workstation. It is thus highly efficient with respect to the required computational resources and does not introduce additional delays in the modeling process. Modifications or extensions in the underlying codebase require only adjusting the general configuration that describes which elements should be exposed. The generation process itself remains unaffected, enabling rapid integration of new features while preserving maintainability. This is particularly relevant as it enables the integration into Python without explicit knowledge of the underlying binding framework.

In future work, we will consider how C++ model and Python binding code could simultaneously be generated by leveraging the already existing C++ code together with user-defined input on model generalizations or extensions, e.g., through the inclusion of vector-borne diseases.

\begin{acks}
This work was supported by the Initiative and Networking Fund of the Helmholtz Association (grant agreement number KA1-Co-08, Project LOKI-Pandemics) and by the German Federal Ministry for Digital and Transport under grant agreement FKZ19F2211A (Project PANDEMOS). It was furthermore supported by the German Federal Ministry of Education and Research and the German Federal Ministry of Research, Technology and Space under grant agreement 031L0325A (Project TwinChain) and 031L0324E (Project EPISERVE) and the Deutsche Forschungsgemeinschaft (DFG, German Research Foundation) (grant agreement 528702961). TwinChain and EPISERVE are part of the Modeling Network for Severe Infectious Diseases (MONID).
\end{acks}

\bibliographystyle{ACM-Reference-Format}
\bibliography{literature}

\section*{Appendix}

\subsection*{Binding text for the simulation function}
\label{appendix: Structure to generate binding string for simulate function}
\begin{lstlisting}[caption={\textbf{Structure to generate binding string to replace the \$simulate placeholder.}},label={code:placeholder_function}]
def simulation(intermed_repr: IntermediateRepresentation) -> str:
    """ Generate the code for the Simulation class.
    Not used by every model.

    :param intermed_repr: Dataclass holding the model features.
    :returns: Formatted string representing a part of the bindings.

    """
    namespace = intermed_repr.namespace

    if intermed_repr.simulation is True:

        simulation = "&" + namespace + "simulate<" + \
            ScalarType(intermed_repr)+">"
        flow_simulation = "&" + namespace + \
            "flow_simulation<"+ScalarType(intermed_repr)+">"
        bind_simulation_class_string = "" + namespace + "Simulation"
        bind_flowsimulation_class_string = "" + namespace + "FlowSimulation"
        bind_simulation_string = ""
    else:

        simulation = "&mio::simulate<"+ScalarType(intermed_repr) + \
            "," + namespace + "Model<"+ScalarType(intermed_repr)+">>"
        flow_simulation = "&mio::flow_simulation<" + \
            namespace + "Model<"+ScalarType(intermed_repr)+">>"
        bind_simulation_class_string = "mio::Simulation"
        bind_flowsimulation_class_string = "mio::FlowSimulation"
        bind_simulation_string = "" + \
            ScalarType(intermed_repr)+"," + namespace + \
            "Model<"+ScalarType(intermed_repr)+">"

    sub_string = ""

    sub_string += (
        "pymio::bind_Simulation<{b_sim}<{sub}>>(m, \"Simulation\");\n\n\t"

        'm.def(\n\t\t"simulate", {sim},\n\t\t'
        'py::arg("t0"), py::arg("tmax"), py::arg("dt"), py::arg("model"), py::arg("integrator") = py::none(),\n\t\t'
        '"Simulate a {namespace} from t0 to tmax."\n\t\t'
        ');\n\n\t'

        "pymio::bind_ModelNode<{namespace}Model<" +
        ScalarType(intermed_repr)+">>(m, \"ModelNode\");\n\t"
        "pymio::bind_SimulationNode<{namespace}Simulation<>>(m, \"SimulationNode\");\n\t"
        "pymio::bind_ModelGraph<{namespace}Model<" +
        ScalarType(intermed_repr)+">>(m, \"ModelGraph\");\n\t"
        "pymio::bind_MobilityGraph<{namespace}Simulation<>>(m, \"MobilityGraph\");\n\t"
        "pymio::bind_GraphSimulation<mio::Graph<mio::SimulationNode<{b_sim}<>>, mio::MobilityEdge<" +
        ScalarType(intermed_repr)+">>>(m, \"MobilitySimulation\");\n\n\t"

    ).format(
        namespace=intermed_repr.namespace,
        sim=simulation,
        b_sim=bind_simulation_class_string,
        sub=bind_simulation_string
    )

    if intermed_repr.is_flowmodel is True:

        sub_string += (
            "pymio::bind_Flow_Simulation<{b_sim}<"+ScalarType(intermed_repr) +
            ", mio::FlowSimulation<{sub}>>>(m, \"FlowSimulation\");\n\n\t"

            'm.def(\n\t\t"simulate_flows", {flow_sim},\n\t\t'
            'py::arg("t0"), py::arg("tmax"), py::arg("dt"), py::arg("model"), py::arg("integrator") = py::none(),\n\t\t'
            '"Simulate a {namespace} with flows from t0 to tmax."\n\t\t'
            ');\n\n\t'

            "pymio::bind_SimulationNode<{namespace}FlowSimulation<>>(m, \"SimulationNode\");\n\t"
            "pymio::bind_MobilityGraph<{namespace}FlowSimulation<>>(m, \"MobilityGraph\");\n\t"
            "pymio::bind_GraphSimulation<mio::Graph<mio::SimulationNode<{b_flowsim}<>>, mio::MobilityEdge<" +
            ScalarType(intermed_repr)+">>>(m, \"MobilitySimulation\");\n\t"

        ).format(
            namespace=intermed_repr.namespace,
            flow_sim=flow_simulation,
            b_sim=bind_simulation_class_string,
            b_flowsim=bind_flowsimulation_class_string,
            sub=bind_simulation_string
        )
    return sub_string
\end{lstlisting}

\subsection*{Parameters, explanations, and average value for sensitivity analysis}

\begin{table}[H]
    \centering
\def\arraystretch{1.25}
\begin{tabular}{ccc}
      Parameter & Explanation & Average value \\
\hline \hline
    $T_{E}$ & Time in exposed state &  $3.335$\\
     $T_{I_{NS}}$ & Time in infectious, nonsymptomatic state & $2.58916$\\
      $T_{I_{Sy}}$ & Time in infectious, symptomatic state & $6.94547$\\
    $T_{I_{Sev}}$ & Time in severe state & $7.28196$\\
     $T_{I_{Cr}}$ & Time in critical state & $13.066$ \\
     \midrule
 $\rho$ & Transmission probability on contact & $0.07333$\\
 $\xi_{I_{NS}}$ & Relative transmission from infectious, nonsymptomatic & 1.0\\
 $\xi_{I_{Sy}}$ & Relative transmission from infectious, symptomatic (nonisolation share) & 0.25\\
\midrule
   1-$\mu_{I_{NS}}^{I_{Sy}}$ & Share of recovered per nonsymptomatic, infectious & $ 0.2069$\\
     $\mu_{I_{Sy}}^{I_{Sev}}$ & Share of severe per symptomatic, infectious & $0.07864$\\
     $\mu_{I_{Sev}}^{I_{Cr}}$ & Share of critical per severe & $0.17318$\\
     $\mu_{I_{Cr}}^{D}$ & Share of dead per critical & $0.21718$\\   
     \midrule
\bottomrule
\end{tabular}
\caption{\textbf{Parameters for wild-type SARS-CoV\protect\nobreakdash-2 based on~\cite{schmidt_gnn_2025}.}}
\label{tab:COVID-19parameters}
\end{table}

\end{document}